\documentclass[journal,final,twocolumn,10pt,twoside]{IEEEtranTCOM}

\IEEEoverridecommandlockouts 

\ifCLASSINFOpdf

\else
\fi
\usepackage{mathtools}
\usepackage{amsmath}
\usepackage{amssymb}
\usepackage{cite}
\usepackage{graphicx}
\usepackage{multirow}
\usepackage{siunitx}
\usepackage{eufrak}
\usepackage{yfonts}
\usepackage{lipsum}
\usepackage{cuted}
\usepackage{multirow}
\usepackage{multicol}
\usepackage{siunitx}
\usepackage{graphicx}
\usepackage{graphicx}
\usepackage{slashbox}
\usepackage{color}
\usepackage{xcolor}
\DeclareMathOperator{\E}{\mathbb{E}}
\begin{document}
\title{Statistical Studies of Fading in Underwater Wireless Optical Channels in the Presence of Air Bubble, Temperature, and Salinity Random Variations\\ (Long Version)} %
\author{Mohammad Vahid Jamali,
        Ali Mirani, Alireza Parsay, Bahman Abolhassani, Pooya Nabavi, Ata Chizari,\\ Pirazh Khorramshahi, Sajjad Abdollahramezani, and~Jawad~A.~Salehi,~\IEEEmembership{\normalsize Fellow,~IEEE 
                \vspace{-0.3in}}
\thanks{Mohammad Vahid Jamali is with the Electrical Engineering and Computer Science Department, University of Michigan, Ann Arbor, MI, USA (e-mail: mvjamali@umich.edu). Ali Mirani, Alireza Parsay, and Jawad A. Salehi are with the Optical Networks Research Laboratory (ONRL), Department of Electrical Engineering, Sharif University of Technology, Tehran, Iran (e-mail: alimirany@gmail.com, parsay\_alireza@ee.sharif.edu, and jasalehi@sharif.edu).
Bahman Abolhassani is with the Department of Electrical and Computer Engineering, Ohio State University, Columbus, OH, USA (e-mail: abolhassani.2@osu.edu). 
Pooya Nabavi is with the Department of Electrical and Computer Engineering, Rice University, Houston, TX, USA (e-mail: pooya.nabavi@gmail.com). Ata Chizari is with the Biomedical Photonic Imaging Group, Faculty of Science and Technology, University of Twente, Enschede, Netherlands (e-mail: a.chizari@utwente.nl).
 Pirazh Khorramshahi is with the Department of Electrical and Computer Engineering, University of Maryland, College Park, MD, USA (e-mail: pkhorram@umd.edu). And Sajjad Abdollahramezani is with the School of Electrical and Computer Engineering, Georgia Institute of Technology, Atlanta, GA, USA (e-mail: ramezani@gatech.edu).
        Part of this paper was presented in the 4th Iran Workshop on Communication and Information Theory, IWCIT'2016, Tehran, Iran.}}

\maketitle
\begin{abstract}
Optical signal propagation through underwater channels is affected by three main degrading phenomena, namely absorption, scattering, and fading. In this paper, we experimentally study the statistical distribution of intensity fluctuations in underwater wireless optical channels with random temperature and salinity variations as well as the presence of air bubbles. In particular, we define different scenarios to produce random fluctuations on the water refractive index across the propagation path, and then examine the accuracy of various statistical distributions in terms of their goodness of fit to the experimental data. We also obtain the channel coherence time to address the average period of fading temporal variations. The scenarios under consideration cover a wide range of scintillation index from weak to strong turbulence. Moreover, the effects of beam-collimator at the transmitter side and aperture averaging lens at the receiver side are experimentally investigated. We show that the use of a transmitter beam-collimator and/or a receiver aperture averaging lens suits single-lobe distributions such that the generalized Gamma and exponentiated Weibull distributions can excellently match the histograms of the acquired data. Our experimental results further reveal that the channel coherence time is on the order of $10^{-3}$ seconds and larger which implies to the slow fading turbulent channels.
\end{abstract}
\IEEEpeerreviewmaketitle
\section{Introduction}
\IEEEPARstart{U}{nderwater} wireless optical communication (UWOC) is becoming a dominant solution for high-throughput and large-data underwater communications thanks to its tremendous advantages, such as higher bandwidth, lower time latency, and better security, compared to the traditional acoustic communication systems. However, the UWOC channel severe impairments including absorption, scattering, and turbulence-induced fading hamper on the signal propagation and limit the viable communication range to typically less than $100$ \si{m} \cite{akhoundi2016cellular,kaushal2016underwater}. Therefore, removing this impediment and enabling UWOC for longer link ranges demand, firstly, comprehensive studies on these degrading effects and accurate channel modeling, and then designing UWOC systems that are robust with respect to these impairments and can reliably communicate over longer distances.

In the past few years, intensive research activities have been carried out to design more reliable and smart UWOC systems that can alleviate one or more of the channel main impairments. For example, in \cite{jamali2015performanceMIMO,jamali2015ber,jamali2017mimovlc} multiple-input multiple-output (MIMO) transmission, with respect to the Monte Carlo (MC)-based fading-free impulse response modeling proposed in \cite{tang2014impulse,cox2012simulation,cox2014simulating,gabriel2013monte} and lognormal fading model for weak turbulence \cite{andrews2001laser,gerccekciouglu2014bit}, has been employed for both diffusive light emitting diode (LED)-based and collimated laser-based UWOC links to mitigate turbulence-induced fading. Moreover, relying on the distance-dependency of all of the channel disturbing effects, the beneficial application of multi-hop serial relaying in alleviating all of the channel impairments has been explored in \cite{jamali2016performance,jamali2016relay}; the results show remarkable increase of the viable communication range  for both point-to-point and optical code division multiple access (OCDMA)-based UWOC systems. 

It is well understood that accurate channel modeling plays a key role on the precise evaluation of the performance of UWOC systems. In this context, while absorption and scattering are well studied for different water types and can be modeled using MC numerical simulations like \cite{tang2014impulse,cox2012simulation,cox2014simulating,zhang2016impulse}, the literature still lacks a comprehensive statistical study over turbulence-induced fading in UWOC channels which is the main scope of this paper.

In contrast to the underwater acoustic and radio frequency (RF) communications where multipath is the main source for signal fading, fading in wireless optical communications occurs as a result of random variations in the refractive index of the conveying medium which is referred to as turbulence. Optical turbulence in UWOC channels is mainly due to the random variations of the water temperature and salinity \cite{korotkova2012light,tang2013temporal,thorpe2007introduction} while in free-space optics (FSO) random variations of the atmosphere pressure and temperature are its main reasons \cite{andrews2001laser,zhu2002free,jazayerifar2006atmospheric}. In addition to temperature- and salinity-induced turbulence, air bubbles can also cause random fluctuations on the received optical signal through the UWOC channel \cite{simpson2009spatial}. This also has some physical interpretations, e.g., when a diver is sending the information bits using optical signals, the surrounding air bubbles produced by the diver may cause fluctuations on the optical signal. In \cite{jamali2016statistical}, the random fluctuations of the received optical signal through the UWOC channel with air bubbles has been experimentally studied and the accuracy of some of the most known statistical distributions in predicting the intensity random fluctuations has been examined. Subsequently, in a recent work \cite{oubei2017performance}, the performance of UWOC systems in the presence of different air bubble populations has experimentally been evaluated. 

As explained before, in a typical UWOC channel, temperature- and/or salinity-induced turbulence is the main cause of fading on the optical signal. However, the accurate statistical distributions for such turbulence-induced fading in UWOC channels are not yet well investigated. In this context, the authors in \cite{bernotas2015probability,bernotas2016probability,oubei2017simple} have experimented statistical distributions of UWOC channels with temperature gradient and in the absence of salinity random variations, however in very weak underwater turbulence regions characterized by the scintillation index values much less than unity ($\sigma^2_I\ll 1$). 
 In this paper, we run various scenarios to produce air bubbles and also induce temperature and salinity random variations across the optical beam propagation. Our scenarios cover a wide range of scintillation index values from weak to strong turbulence. We then examine the accuracy of different statistical distributions in terms of their goodness of fit and comment on the applicability of these distributions in each range of the scintillation index. We further obtain the fading coherence time to confirm that the UWOC turbulent channels are usually under slow fading.
 
 The remainder of the paper is organized as follows. In Section II,
we briefly review the preliminaries pertaining the UWOC channel with emphasis on the turbulence-induced fading. In Section III, we
define the goodness of fit metric, and explain how the channel coherence time can experimentally be calculated to ascertain the slow fading nature of UWOC turbulent channels. In Section IV, we describe the statistical distributions under consideration with insights to their application in general wireless optical channels.
 Section V explains the system model and experimental set-up,
 Section VI presents the comprehensive experimental results for various scenarios,
 and Section VII concludes the paper and highlights some relevant future research directions.
 \section{UWOC Channel Overview}
 The channel modeling plays a critical role in ascertaining the performance of UWOC systems and predicting the outcomes of different techniques. Comprehensive studies on the channel modeling have revealed that the propagation of light under water is affected by three major degrading effects, i.e., absorption, scattering, and turbulence \cite{mobley1994light,korotkova2012light}. In fact, during the propagation of photons through water, they may interact with water molecules and particles. In this case, the energy of each photon may be lost thermally, that is named absorption process and is characterized by absorption coefficient $a(\lambda)$, where $\lambda$ is the wavelength. And the direction of each photon may be altered which is defined as scattering process and is determined by scattering coefficient $b(\lambda)$. Then the total energy loss of non-scattered light is described by extinction coefficient $c(\lambda)=a(\lambda)+b(\lambda)$ \cite{mobley1994light}.
 
 Many invaluable researches have been carried out in the past few years 
concerning on these two fundamental effects of UWOC channels, and thorough channel modeling based on MC numerical method with respect to absorption and scattering effects have been reported in several recent work \cite{cox2012simulation,cox2014simulating,tang2014impulse,jamali2017mimovlc}. Subsequently, some other recent research activities have focused on proposing closed-form expressions for the characterization of the fading-free impulse response (FFIR) of both point-to-point and MIMO UWOC links based on MC numerical simulations \cite{tang2014impulse,zhang2016impulse}.
Although the above-mentioned FFIR takes into account both absorption and scattering effects, accurate and complete characterization of the UWOC channel impulse response requires turbulence-induced fading consideration as well. A fading coefficient can be multiplied by the channel FFIR to include the slow fading nature of UWOC channels \cite{tang2013temporal,andrews2001laser}. Although the recent mathematical works on the performance evaluation of UWOC systems \cite{yi2015underwater,gerccekciouglu2014bit,jamali2016relay,jamali2015performanceMIMO}, inspired by the behaviour of optical turbulence in atmosphere, have considered lognormal probability density function (PDF), the accurate characterization of fading coefficients' PDF demands a more specific investigation which will be experimentally carried out in this paper.
 
In order to measure the fading strength, it is common in the literature to define the scintillation index of a propagating light wave as \cite{andrews2005laser,korotkova2012light}
  \begin{align} \label{S.I.}
 {\sigma }^2_I\left(r,d_0,\lambda \right)=\frac{\E\left[I^2(r,d_0,\lambda )\right] -{{\E}^2\left[I\left(r,d_0,\lambda \right)\right]}}{{{\E}^2\left[I\left(r,d_0,\lambda \right)\right]}},
 \end{align}
where $I(r,d_0,\lambda )$ is the instantaneous intensity at a point with position vector $\left(r,d_0\right)=(x,y,d_0)$, $d_0$ is the link length, and $\E\left[I\right]$ denotes the expected value of the random variable (RV) $I$.
There are lots of theoretical research activities toward mathematical investigation of the UWOC channel turbulence and how the scintillation index varies with the water specific parameters of turbulence \cite{korotkova2012light,yi2015underwater,gerccekciouglu2014bit,ata2014scintillations}. However, non of these prior works have specifically focused on the characterization of the UWOC fading statistical distribution which is a necessary task from the communication engineering point of view; this important study shapes the main body of our paper.
We should emphasize that each statistical distribution has a set of constant parameters which directly relate to the scintillation index value; therefore, as it will be clarified in Section IV, these parameters vary with the fading strength.
\section{Evaluation Metrics}
 \subsection{Goodness of Fit}
In order to test a distribution's fitness and evaluate the accordance of different statistical distributions with the experimental data, we use goodness of fit metric, also known as $R^2$ measure,
 \begin{align}\label{R^2}
 R^2=1-\frac{SS_{\rm reg}}{SS_{\rm tot}},
 \end{align}
 in which $SS_{\rm reg}$ is the sum of the square errors of the statistical distribution under consideration, i.e., $SS_{\rm reg}=\sum_{i=1}^{M}\left(f_{m,i}-f_{p,i}\right)^2$, where $M$ is the number of bins of the acquired data histogram, $f_{m,i}$ and $f_{p,i}$ are respectively the measured and predicted probability values for a given received intensity level corresponding to the $i$th bin. And $SS_{\rm tot}$ is the sum of the squares of distances between the measured points and their mean, i.e., $SS_{\rm tot}=\sum_{i=1}^{M}\left(f_{m,i}-\bar{f}\right)^2$, where $\bar{f}=\sum_{i=1}^{M}f_{m,i}/M$. Clearly, as the value of the $R^2$ measure for a given distribution approaches its maximum (i.e., $1$), the distribution better fits the measured data.
 \subsection{Coherence Time}
  In order to verify the slow fading nature of UWOC channels and corroborate the theoretical achievements in \cite{tang2013temporal}, we experimentally measure the channel coherence time for different channel conditions. To do so, we measure the time in which the temporal covariance coefficient of irradiance, defined as Eq. \eqref{NTC} \cite{tang2013temporal,andrews2005laser}, remains above a certain threshold, e.g., a $-3$ \si{dB} threshold.
  \begin{align}\label{NTC}
 b_{\tau,I}(d_0,\tau)=\frac{B_{\tau,I}(d_0,\tau)}{B_{\tau,I}(d_0,0)},
  \end{align}
  where $B_{\tau,I}(d_0,\tau)\stackrel{\triangle}{=}B_{I}(\boldsymbol{r_1},d_0,t_1;\boldsymbol{r_2},d_0,t_2)$ with $\boldsymbol{r_1}=\boldsymbol{r_2}$. Furthermore, $\tau=t_1-t_2$, and $B_{I}(\boldsymbol{r_1},d_0,t_1;\boldsymbol{r_2},d_0,t_2)$ is defined as the covariance of irradiance for two points $\boldsymbol{r_1}$ and $\boldsymbol{r_2}$ at different time instants $t_1$ and $t_2$ \cite{andrews2005laser}.
 \section{The Probability Density Functions Under Consideration}
 In this section, we overview the statistical distributions considered in this paper. It is worth mentioning that each statistical distribution contains a set of parameters which can be obtained such that they simultaneously satisfy two essential criteria.
{The first criteria is known as the fading normalization equality which means that the fading coefficient is normalized to emphasize that it neither amplifies nor attenuates the average power, i.e., $\E[{\tilde{h}}]=1$. To do so during our experiments, we normalize the received vector from each channel realization, corresponding to the received optical power from a specific turbulent UWOC channel at each scenario, to its mean to make sure that the mean of the new vector is equal to one.}
 Additionally, as the second criteria, we run an optimization procedure to obtain the distribution parameters such that they result into the best fit to the experimental data in terms of the goodness of fit. 

  \subsection{Lognormal Distribution}
Lognormal distribution is mainly used in the literature to describe the fluctuations induced by weak atmospheric turbulence, characterized by $\sigma^2_I<1$. In this case, the channel fading coefficient $\tilde{h}$ has the PDF of
     \begin{equation} \label{log-normal}
    f_{\tilde{h}}({\tilde{h}})=\frac{1}{2\tilde{h} \sqrt{2\pi {\sigma }^2_X}}{\rm exp}\left(-\frac{{\left({{\rm ln}(\tilde{h})\ }-2{\mu }_X\right)}^2}{8{\sigma }^2_X}\right),
     \end{equation}
    where $\mu_X$ and $\sigma^2_X$ are respectively the mean and variance of the Gaussian-distributed fading log-amplitude factor defined as $X=1/2\ln(\tilde{h})$ \cite{andrews2001laser}.
      The fading normalization equality for the lognormal distribution leads to ${{\mu }_X=-\sigma }^2_X$ \cite{andrews2001laser}. Therefore, lognormal distribution is a function of a single parameter ${\sigma }^2_X$ which is related to the scintillation index value as ${\sigma }^2_X=0.25\ln(1+\sigma^2_I)$ \cite{jamali2015performanceMIMO}.
  \subsection{Gamma Distribution}
  The Gamma distribution with shape parameter $k$ and scale parameter $\theta$ is expressed as
\begin{align}\label{gamma}
f_{\tilde{h}}(\tilde{h})=\frac{1}{\Gamma(k)\theta^k}\tilde{h}^{k-1}\exp\left(-{\tilde{h}}/{\theta}\right),
\end{align}
where $\Gamma(k)$ is the Gamma function. For the Gamma distribution we note that $\E[\tilde{h}]=k\theta=1$ and $\E[\tilde{h}^2]=k(k+1)\theta^2$. Therefore, the Gamma distribution parameters are simply related to the scintillation index value as $\theta=1/k=\sigma^2_I$. 
\subsection{K Distribution}
This distribution is mainly used in the FSO literature for strong atmospheric turbulence in which $\sigma^2_I\geq 1$. The PDF of K distribution is given by \cite{andrews2001laser}
       \begin{align}\label{K distribution}
       f_{\tilde{h}}(\tilde{h})=\frac{2\alpha}{\Gamma(\alpha)}\left(\alpha\tilde{h}\right)^{(\alpha-1)/2}K_{\alpha-1}\left(2\sqrt{\alpha\tilde{h}}~\!\right),
       \end{align}
in which $K_p(x)$ is the $p$th-order modified Bessel function of the second kind and $\alpha$ is a positive parameter which relates to the scintillation index value as $\alpha=2/(\sigma^2_I-1)$. It is worth noting that the PDF of K distribution expressed in Eq. \eqref{K distribution} already satisfies the fading normalization criteria, $\E[\tilde{h}]=1$.
  \subsection{Weibull Distribution}
  The Weibull and exponentiated Weibull distributions were used in \cite{barrios2013exponentiated} to excellently describe the atmospheric turbulence in a wide range of scintillation index values. In this paper, we evaluate the accordance of these two distributions as good candidates for predicting the fluctuations in the received optical signal through a turbulent UWOC channel. The Weibull PDF is defined as \cite{bernotas2016probability}
  \begin{align}\label{Weibull}
  f_{\tilde{h}}(\tilde{h})=\frac{\beta}{\eta}\left({\tilde{h}}/{\eta}\right)^{\beta-1}\exp\left(-\left({\tilde{h}}/{\eta}\right)^{\beta}\right).
  \end{align}
For the Weibull distribution we find that $\E[\tilde{h}]=\eta\Gamma\left(1+1/\beta\right)$ and $\E[\tilde{h}^2]=\eta^2\Gamma\left(1+2/\beta\right)$. Therefore, the fading normalization equality results into the relation $\eta=1/\Gamma\left(1+1/\beta\right)$ between $\eta$ and $\beta$. Moreover, the scintillation index value can be obtained from the value of the parameter $\beta$ as $\sigma^2_I=\left[\Gamma\left(1+2/\beta\right)/\Gamma^2\left(1+1/\beta\right)\right]-1$.
  \subsection{Exponentiated Weibull Distribution}
  The PDF of a RV described by the exponentiated Weibull distribution with parameters $\alpha$, $\beta$, and $\eta$ is given by
    \begin{align}\label{exponentiated Weibull}
    f_{\tilde{h}}(\tilde{h})=&\frac{\alpha\beta}{\eta}\left({\tilde{h}}/{\eta}\right)^{\beta-1}\exp\left(-\left({\tilde{h}}/{\eta}\right)^{\beta}\right)\nonumber\\
    &\times\left[1-\exp\left(-\left({\tilde{h}}/{\eta}\right)^{\beta}\right)\right]^{\alpha-1}.
    \end{align}
The $n$th moment of the exponentiated Weibull PDF can be calculated as \cite{nadarajah2005moments,barrios2013exponentiated2}
\begin{align}\label{nth_moment}
\E\big[\tilde{h}^n\big]=\alpha\eta^n\Gamma\left(1+n/\beta\right)g_n(\alpha,\beta),
\end{align}
where $g_n(\alpha,\beta)$ is a series defined as
\begin{align}\label{g_n}
g_n(\alpha,\beta)=\sum_{i=0}^{\infty}\frac{(-1)^i\Gamma(\alpha)}{i!(i+1)^{1+n/\beta}\Gamma(\alpha-i)}.
\end{align}
As it has been explained in \cite{bernotas2016probability,barrios2013exponentiated2}, Eq. \eqref{g_n} very quickly converges and considering the first $10$ terms suffices to acceptably approximate the infinite series of $g_n(\alpha,\beta)$. Based on Eq. \eqref{nth_moment}, the first moment for satisfying the fading normalization criteria and the second moment for the calculation of the scintillation index can be obtained. The resulted scintillation index formula for the exponentiated Weibull distribution is expressed as
\begin{align}\label{sI2_expwieb}
\sigma^2_I=\frac{\Gamma(1+2/\beta)g_2(\alpha,\beta)}{\alpha\left[\Gamma(1+1/\beta)g_1(\alpha,\beta)\right]^2}-1.
\end{align}


  \subsection{Gamma-Gamma Distribution}
  Such a statistical model, which factorizes the irradiance as the product of two independent random
    processes each with a Gamma PDF, has been widely employed in the FSO literature to model a wide range of atmospheric turbulence. The PDF of Gamma-Gamma distribution is expressed as \cite{andrews2001laser};
   \begin{align}\label{GG distribution}
     f_{\tilde{h}}(\tilde{h})=\frac{2(\alpha\beta)^{(\alpha+\beta)/2}}{\Gamma(\alpha)\Gamma(\beta)}(\tilde{h})^{\frac{(\alpha+\beta)}{2}-1}K_{\alpha-\beta}\left(2\sqrt{\alpha\beta\tilde{h}}~\!\right),
   \end{align}
    in which $\alpha$ and $\beta$ are parameters related to the effective atmospheric conditions. The PDF in Eq. \eqref{GG distribution} is normalized so that it satisfies the equality $\E[\tilde{h}]=1$. Furthermore, the scintillation index for Gamma-Gamma distribution is given by $\sigma^2_I=1/\alpha+1/\beta+1/\alpha\beta$ \cite{andrews2001laser}.
  \subsection{Generalized Gamma Distribution}
  In order to simultaneously cover the features of some of the statistical distributions, in this paper, we also evaluate the accordance of the generalized Gamma distribution, defined as
  \begin{align}\label{GGD}
  f_{\tilde{h}}(\tilde{h};a,d,p)=\frac{p}{a^d\Gamma(d/p)}\tilde{h}^{d-1}\exp\left(-({\tilde{h}}/{a})^{p}\right),
  \end{align}
  in predicting the underwater turbulence-induced fading. In fact, the generalized Gamma distribution is a general case of some of the important statistical distributions in describing the optical turbulence. Particularly, when $d=p$ it reduces to the Weibull distribution, for $p=1$ it is equivalent to the Gamma distribution, and if $d=p=1$, the generalized Gamma distribution becomes the exponential distribution \cite{parr1965method}. A simple mathematical manipulation for the derivation of the moments shows that for the generalized Gamma distribution $\E[\tilde{h}]=a\Gamma(d/p+1/p)/\Gamma(d/p)$ and $\E[\tilde{h}^2]=a^2\Gamma(d/p+2/p)/\Gamma(d/p)$. In this case, the scintillation index can be expressed as
  \begin{align}\label{sigma2_I}
  \sigma^2_I=\frac{\Gamma(d/p)\Gamma\left(\frac{d+2}{p}\right)}{\Gamma^2\left(\frac{d+1}{p}\right)}-1.
  \end{align}
  Similar to the previous statistical distributions, in order to obtain the distribution parameters, namely $a$, $d$, and $p$, we use an optimization procedure to simultaneously satisfy the equality $\E[\tilde{h}]=a\Gamma(d/p+1/p)/\Gamma(d/p)=1$ and obtain the best goodness of fit.
 
 \section{Experimental Setup}
\begin{figure*}
 \centering
 \includegraphics[width=7in]{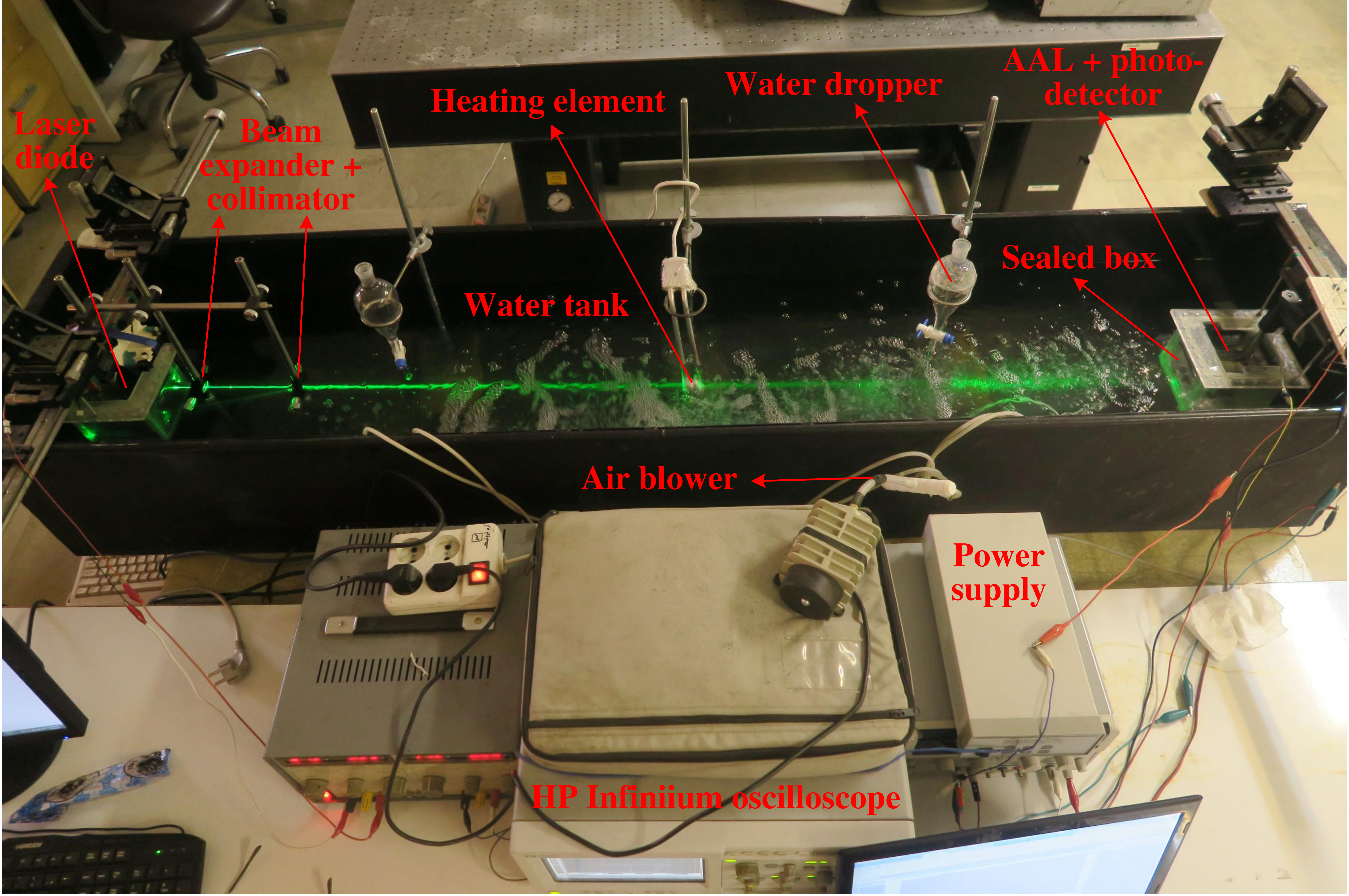}
 \caption{Experimental setup for the characterization of turbulence-induced fading in UWOC channels.}
 \label{fig:set_up}
 \vspace{-0.1in}
 \end{figure*}
 Fig. 1 shows our experimental setup for collecting the received intensity samples. As it is shown, our experimental setup is implemented in a black-colored water tank with dimensions $30\times 40\times 200$ \si{cm^3}.
 At the transmitter side, a $532$ green laser diode with maximum output power of $50$ \si{mW} is driven by a MOS transistor to ensure a constant optical irradiance. Furthermore, to collimate and expand the transmitter beam, we have designed a beam-collimator (BC) using a plano-concave lens of focal length $f=-30$ \si{mm} followed by a plano-convex lens with $f=200$ \si{mm}. The diameter of both of the lenses of the BC is $D_{0,t}=25$ \si{mm}. On the other hand, we use an ultraviolate-visible photodetector at the receiver side to capture the optical irradiance. The photodetector converts the received optical power $p(t)$ to the electrical current $i(t)$ according to the relation $i(t)=\eta e p(t)/hf$, where $\eta$ is the photodetector's quantum efficiency, $e$ is the electron charge, $h$ is the Planck's constant, and $f$ is the light frequency. Therefore, the photodetector's output current is proportional to the received optical power and hence is a good representative of the received irradiance. We further amplify the photo-detected current and then sample and monitor the amplified signal by an HP Infiniium oscilloscope. For each test, we have collected $32768$ samples with the sampling rate of ${5}$ \si{{kSa}/{s}}. Additionally, in order to have a good accommodation with the conventional receivers in wireless optics, communicating through turbulent media, the receiver is equipped with a lens of diameter $D_{0,r}=50$ \si{mm} for the sake of aperture averaging.
 Both the transmitter laser and the receiver photodetector are
 sealed by placing in transparent boxes.

Various scenarios and sets of experiments are performed in this paper. The first scenario deals with the characterization of intensity fluctuation due to the random presence of air bubbles through the propagation path. In this case, in order to generate random air bubbles, a tunable air blower with the maximum blowing capacity of $28$ \si{Litre/m} is employed. The received intensity is sampled for both the absence and presence of the transmitter BC and also the receiver aperture averaging lens (AAL). The second scenario, contains a set of various experiments concentrating on the statistical characterization of the underwater turbulence-induced fading due to the random temperature variations. To produce such random variations on the water temperature, we employ two different mechanisms. In the first method, we use three tunable water droppers to create three independent flow of hot water, with the temperature $T_h=90^{\rm o} \si{C}$, to the cold water of tank with the temperature $T_c=20^{\rm o} \si{C}$. As the second mechanism for generating random temperature variations, we insert three circular heating elements in the propagation path and pass the laser beam through the heaters. Moreover, various mixtures of these two methods for generating temperature variations are also considered in the presence of random air bubbles for the sake of generality.
 In the third set of experiments, in order to investigate the accordance of different statistical distributions in predicting the turbulence-induced fading due to the water random salinity variations, similar to the second scenario, we use three independently tunable water droppers to insert three flow of extremely salty water into the water tank.
 
 \section{Experimental Results}
 In this section, we take into account all of the three different scenarios mentioned in the previous section to experimentally evaluate the validity of various statistical distributions in predicting turbulence-induced fading in UWOC channels under a wide range of channel conditions. For each scenario, we have adjusted the transmitter laser power to ensure that a considerable power reaches the receiver.
  We further numerically calculate the temporal covariance coefficient of the received irradiance for various channel conditions to determine the fading coherence time as an important parameter in the evaluation of the performance of UWOC systems.
  
  {In order to cover a wide range of optical turbulence, we first run an abundant number of experiments each by slightly changing the air bubble population and hot or salty water flow rates, depending on the various scenarios considered in this paper. Then we choose from these many experimental results such that at least contain one from the small $\sigma^2_I$ values, one from the $\sigma^2_I$ values around the averages, and another from the highest possible scintillation index values we can get by tuning the air bubble population or hot/salty water flow rates at the highest values. In what follows, we first report few experimental results, pertaining to the first scenario, performed over the UWOC channels without the transmitter BC and receiver AAL to better elucidate the possibility of two-lobe statistical distributions we proposed recently in \cite{jamali2016statistical}. Then, in order to be commensurate with the practical link configurations, we only focus on the cases where the receiver AAL is employed with/without the transmitter beam collimator.}
 \subsection{Intensity Fluctuation due to the Random Presence of Air Bubbles}        
\begin{figure*}
    \centering
    \includegraphics[width=7in,height=9in]{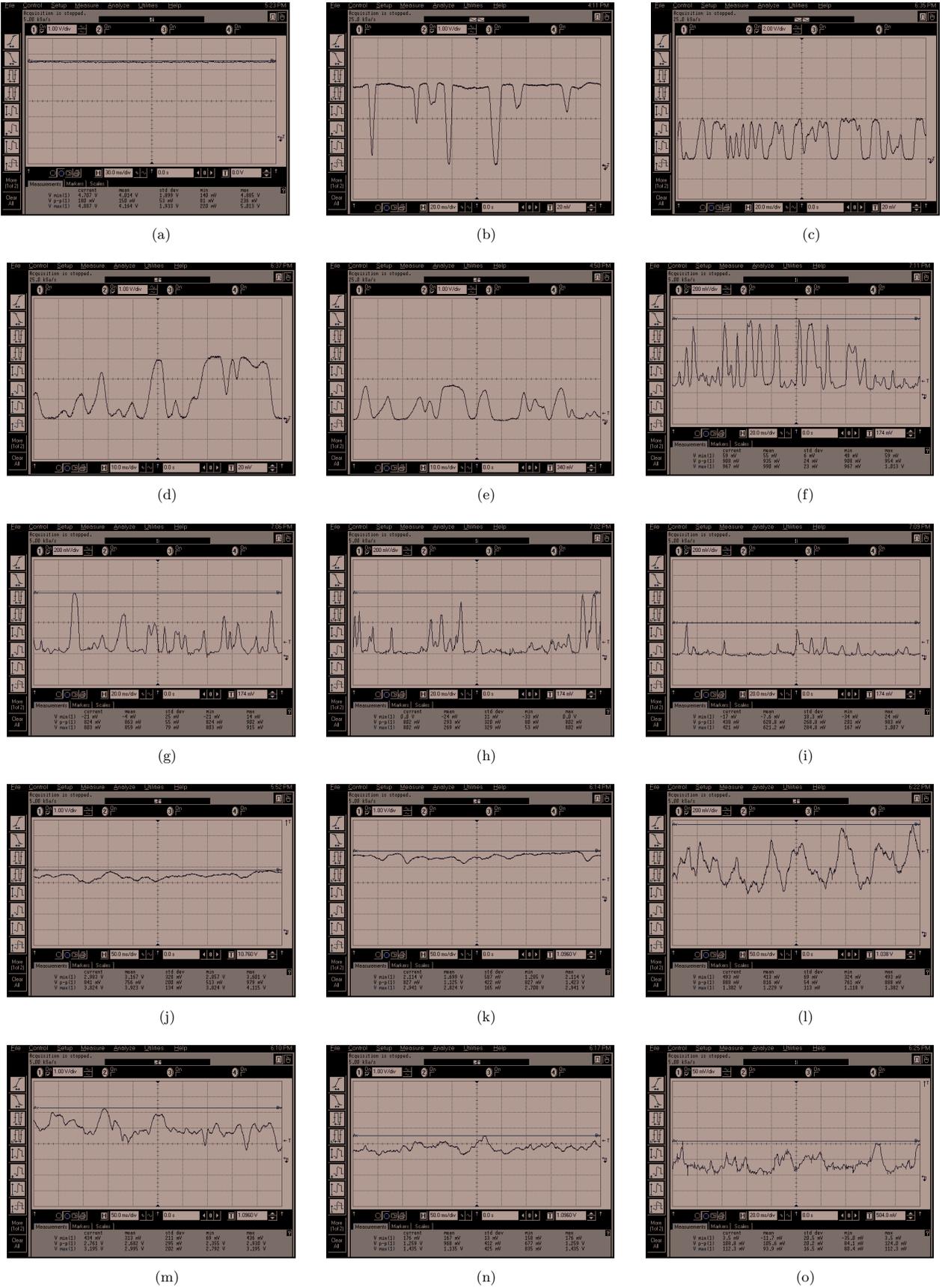}
    \caption{Received optical signal through (a) a fresh water link with $\sigma^2_I=3.65 \times 10^{-5}$; (b) a bubbly water link without the transmitter BC and the receiver AAL, $\mathfrak{B}$, with $\sigma^2_I=0.1014$; (c) a $\mathfrak{B}$ link with $\sigma^2_I=0.2408$; (d) a $\mathfrak{B}$ link with $\sigma^2_I=0.7606$; (e) a $\mathfrak{B}$ link with $\sigma^2_I=3.3415$; (f) a bubbly water link without the transmitter BC and with the receiver AAL, $\mathfrak{B}_{\rm AAL}$, with $\sigma^2_I=0.7148$; (g) a $\mathfrak{B}_{\rm AAL}$ link with $\sigma^2_I=0.9790$; (h) a $\mathfrak{B}_{\rm AAL}$ link with $\sigma^2_I=1.8482$; (i) a $\mathfrak{B}_{\rm AAL}$ link with $\sigma^2_I=2.3285$; (j) a bubbly water link with the transmitter BC and receiver AAL, $\mathfrak{B}^{\rm BC}_{\rm AAL}$, with $\sigma^2_I=0.0047$; (k) a $\mathfrak{B}^{\rm BC}_{\rm AAL}$ link with $\sigma^2_I=0.0144$; (l) a $\mathfrak{B}^{\rm BC}_{\rm AAL}$ link with $\sigma^2_I=0.0628$; (m) a $\mathfrak{B}^{\rm BC}_{\rm AAL}$ link with $\sigma^2_I=0.0986$; (n) a $\mathfrak{B}^{\rm BC}_{\rm AAL}$ link with $\sigma^2_I=0.2082$; and (o) a $\mathfrak{B}^{\rm BC}_{\rm AAL}$ link with $\sigma^2_I=0.4786$.}
    \label{fig_two}
    \end{figure*}

\begin{figure*}
        \centering
        \includegraphics[trim=0cm 0cm 0cm 0.5cm,width=6.8in,height=8.5in,clip]{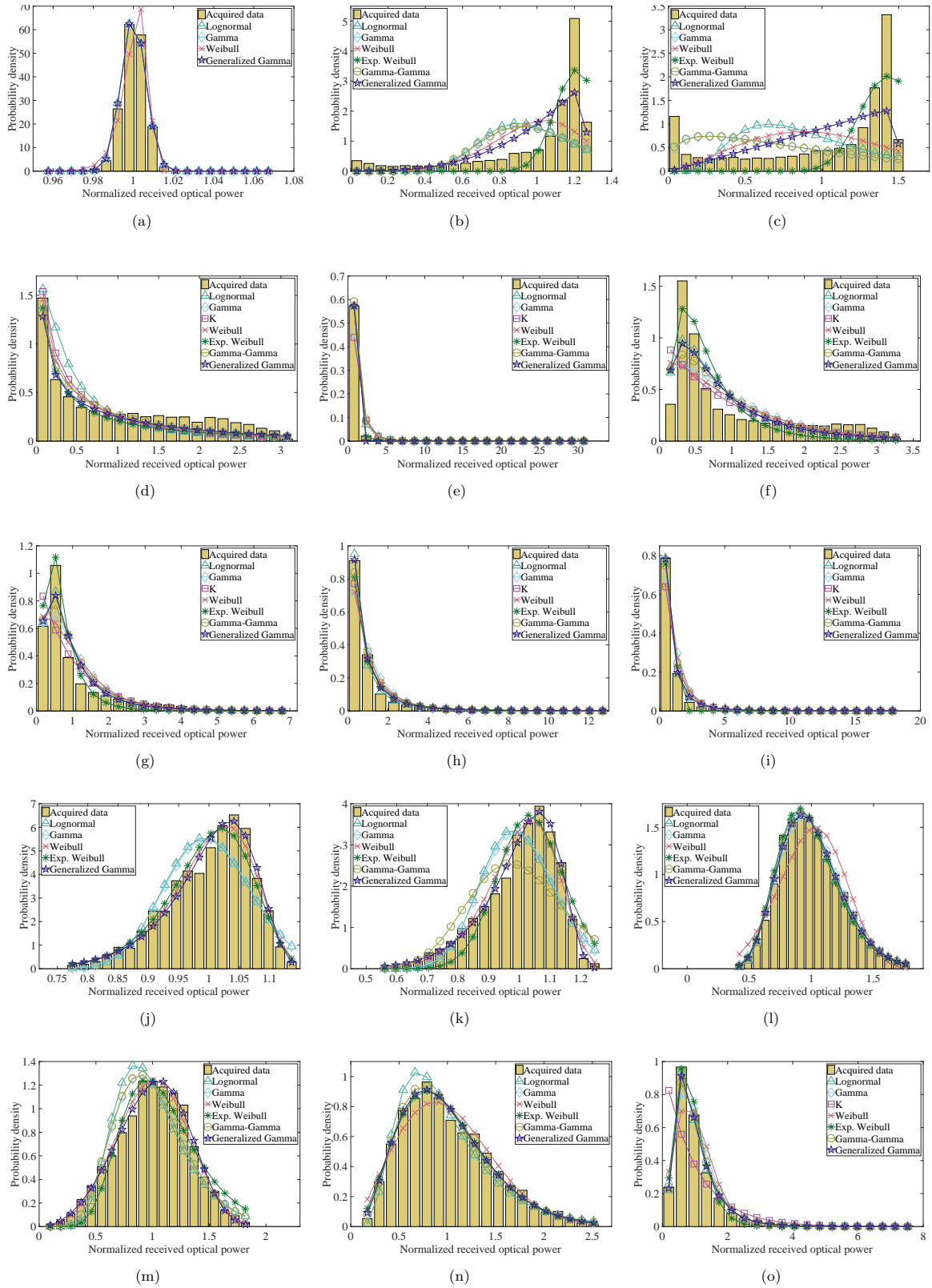}
        \caption{Accordance of different statistical distributions, considered in Sec. IV, with the histograms of the acquired data through various UWOC channel conditions under the random presence of air bubbles, namely (a) a fresh water link with $\sigma^2_I=3.65 \times 10^{-5}$; (b) a bubbly water link without the transmitter BC and the receiver AAL, $\mathfrak{B}$, with $\sigma^2_I=0.1014$; (c) a $\mathfrak{B}$ link with $\sigma^2_I=0.2408$; (d) a $\mathfrak{B}$ link with $\sigma^2_I=0.7606$; (e) a $\mathfrak{B}$ link with $\sigma^2_I=3.3415$; (f) a bubbly water link without the transmitter BC and with the receiver AAL, $\mathfrak{B}_{\rm AAL}$, with $\sigma^2_I=0.7148$; (g) a $\mathfrak{B}_{\rm AAL}$ link with $\sigma^2_I=0.9790$; (h) a $\mathfrak{B}_{\rm AAL}$ link with $\sigma^2_I=1.8482$; (i) a $\mathfrak{B}_{\rm AAL}$ link with $\sigma^2_I=2.3285$; (j) a bubbly water link with the transmitter BC and receiver AAL, $\mathfrak{B}^{\rm BC}_{\rm AAL}$, with $\sigma^2_I=0.0047$; (k) a $\mathfrak{B}^{\rm BC}_{\rm AAL}$ link with $\sigma^2_I=0.0144$; (l) a $\mathfrak{B}^{\rm BC}_{\rm AAL}$ link with $\sigma^2_I=0.0628$; (m) a $\mathfrak{B}^{\rm BC}_{\rm AAL}$ link with $\sigma^2_I=0.0986$; (n) a $\mathfrak{B}^{\rm BC}_{\rm AAL}$ link with $\sigma^2_I=0.2082$; and (o) a $\mathfrak{B}^{\rm BC}_{\rm AAL}$ link with $\sigma^2_I=0.4786$.}
        \label{fig_three}
        \end{figure*}

In this subsection, we present the detailed results for a comprehensive set of experiments performed over UWOC channels under the random presence of air bubbles. Moreover, the effects of beam-collimator (and expander) in the transmitter side and aperture averaging in the receiver side are also investigated. Throughout this subsection, we denote the UWOC channels under air bubble random variations with $\mathfrak{B}$, if the UWOC link is without BC at the transmitter side and AAL at the receiver side. On the other hand, if the UWOC link possesses only the receiver AAL, the scenario is denoted by $\mathfrak{B}_{\rm AAL}$. And if both of them exist in the configuration, we represent the scenario as $\mathfrak{B}^{\rm BC}_{\rm AAL}$. For all of these scenarios we run lots of experiments to cover a wide range of scintillation index, from weak to strong fading.
The temporal domain fluctuations for each of the experiments are shown in Fig. \ref{fig_two}, while the corresponding histograms of the normalized received optical power as well as the fitted statistical distributions are illustrated in Fig. \ref{fig_three}. Moreover, detailed simulation results for the goodness of fit (GoF) values of each of the discussed statistical distributions in Sec. IV and their corresponding constant coefficients are listed in Table I. 

 \begin{table*}
 \vspace{-0.1in}
 		\centering
 		\caption{GoF and the constant parameters of different PDFs for the various channel conditions considered in Scenario I (Sec. VI-A).}
 		\begin{tabular}{||p{0.45in}||p{0.35in}||p{0.6in}||p{0.6in}||p{0.6in}||p{0.6in}||p{0.6in}||p{0.7in}||p{0.6in}||}
 			\hline\hline
 			\! Channel condition \vspace{-2cm} & \vspace{0.25cm} $\sigma^2_{I,{m}}$& 
\vspace{0.1cm} {Log-normal}&
\vspace{0.1cm} {Gamma}&
\vspace{0.1cm} {K dist.}&
\vspace{0.1cm} {Weibull}&
\vspace{0.1cm} $\!\!${Exp. Weibull}&
\vspace{0.1cm} $\!\!${Gamma-Gamma}&
\vspace{0.1cm} $\!\!${Generalized Gamma}\\
 			\cline{3-9}
&& $\begin{matrix} \!{\rm GoF}, \\ \!(\mu_X,\sigma^2_X,\sigma^2_I) \end{matrix}$ &
 $\begin{matrix} {\rm GoF}, \\ (\theta,k,\sigma^2_I) \end{matrix}$ & 
 $\begin{matrix} {\rm GoF}, \\ (\alpha,\sigma^2_I) \end{matrix}$ & 
$\begin{matrix} {\rm GoF}, \\ (\beta,\eta,\sigma^2_I) \end{matrix}$ &
$\begin{matrix} {\rm GoF}, \\ (\alpha,\beta,\eta,\sigma^2_I) \end{matrix}$ & 
 $\begin{matrix} {\rm GoF}, \\ (\alpha,\beta,\sigma^2_I) \end{matrix}$ & 
$\begin{matrix} {\rm GoF}, \\ (a,d,p,\sigma^2_I) \end{matrix}$ 
\\ \hline \hline 			
 			Fresh water &$3.65 \times 10^{-5}$ & $\begin{matrix} \!{0.997}, \\ \!\!\!\!(-9.14\!\times\!\!10^{-6},\\ 9.14\!\times\!\! 10^{-6},\\ 3.65\!\times\!\! 10^{-5}) \end{matrix}$
 			 & $\begin{matrix} \!{0.997}, \\ \!\!\!\!(3.65\!\times\!\!10^{-5},\\ 2.73\!\times\!\! 10^{4},\\ 3.65\!\times\!\! 10^{-5}) \end{matrix}$
 			  & ${\rm Not~Valid}$ & $\begin{matrix} \!{0.952}, \\ \!\!\!\!(189.259,\\ 1.003,\\ 4.55\!\times\!\! 10^{-5}) \end{matrix}$
 			  & ${\rm Not~Valid}$ & ${\rm Not~Valid}$ & $\begin{matrix} \!{0.997}, \\ \!\!\!\!(3.65\!\times\!\!10^{-5},\\ 2.73\!\times\!\! 10^{4},1,\\ 3.65\!\times\!\! 10^{-5}) \end{matrix}$ \\ \hline
 			  
 			  $\mathfrak{B}$ & $0.1014$ & $\!\!\!\!\begin{matrix} 0.0117, \\ (-0.0181,\\ 0.0181,0.0752) \end{matrix}$
 			   			 & $\!\!\!\!\begin{matrix} 0.0753, \\ (0.0750,\!13.333,\\0.0750) \end{matrix}$
 			   			  & ${\rm Not~Valid}$ & $\!\!\!\begin{matrix} {0.2563}, \\ (4.904,1.090,\\ 0.0544) \end{matrix}$
 			   			  & $\begin{matrix} {0.7428}, \\ (6.76,4.82,\\1.02,0.0096) \end{matrix}$ & $\begin{matrix} 0.0660, \\ 
 			   			  (13.9,164.1,\\0.0785) \end{matrix}$
 			   			  & $\!\!\!\!\begin{matrix} {0.6501}, \\ (1.2682,3.847,\\~73.72,0.0415) \end{matrix}$\\ \hline
 			   			  
 $\mathfrak{B}$ & $0.2408$ & $\!\!\!\!\begin{matrix}  -0.5117, \\ (-0.0708,\\ 0.0708,0.3272) \end{matrix}$
 & $\!\!\!\!\begin{matrix}  -0.3510, \\ (0.7320,\!1.3661,\\0.7320) \end{matrix}$
& ${\rm Not~Valid}$ & $\!\!\!\begin{matrix} {-0.2298}, \\ (2.301,1.1288,\\ 0.2124) \end{matrix}$
& $\begin{matrix} {0.4064}, \\ (7.16,3.36,\\1.12,0.0189) \end{matrix}$ & $\begin{matrix} -0.3538, \\ 
(1.35,171.5,\\ 0.7509) \end{matrix}$
 & $\!\!\!\!\begin{matrix} {0.2632}, \\ (\!1.5014,\!2.0374,\\~73.72,0.1209) \end{matrix}$ \\ \hline
 
 $\mathfrak{B}$ & $0.7606$ & $\!\!\!\!\begin{matrix}  0.6121, \\ (-0.4021,\\ 0.4021,3.9942) \end{matrix}$
  & $\!\!\!\!\begin{matrix}  0.9026, \\ (2.044,\!0.4892,\\2.044) \end{matrix}$
 & $\!\!\begin{matrix}   0.8284, \\ (1.39,2.4388) \end{matrix}$ & $\!\!\!\begin{matrix} {0.8736}, \\ (0.693,0.7824,\\ 2.1887) \end{matrix}$
 & $\begin{matrix} {0.8844}, \\ (3.50,0.31,\\0.11,8.5491) \end{matrix}$ & $\begin{matrix} 0.8899, \\ 
 (0.60,171.25,\\ 1.6822) \end{matrix}$
  & $\!\!\!\!\!\!\begin{matrix} {0.9061}, \\ (2.725,\!0.4521,\\~1.3097,0.6075) \end{matrix}$ \\ \hline
  
$\mathfrak{B}$ & $3.3415$ & $\!\!\!\!\begin{matrix}  0.9921, \\ (-0.1906,\\ 0.1906,1.1434) \end{matrix}$
    & $\begin{matrix}  0.9371, \\ (1,1,1) \end{matrix}$
   & $\begin{matrix}   0.9279, \\ (10,1.2) \end{matrix}$ & $\!\!\!\begin{matrix} { 0.9823}, \\ (1.352,1.0908,\\ 0.559) \end{matrix}$
   & $\begin{matrix} {0.9994}, \\ (8,0.61,\\0.11,0.6280) \end{matrix}$ & $\!\!\!\begin{matrix} 0.9856, \\ 
   (1.85,10,0.6946) \end{matrix}$
    &$\!\!\!\!\!\!\begin{matrix} {0.9978}, \\ (1.8076,\!1.280,\\~~\!33.580,\!0.0542) \end{matrix}$ \\ \hline
    
    $\mathfrak{B}_{\rm AAL}$ & $0.7148$ & $\!\!\!\!\begin{matrix}  0.7325, \\ (-0.1889,\\ 0.1889,1.1289) \end{matrix}$
        & $\!\!\!\begin{matrix}  0.5584, \\ (0.735,1.3605,\\0.7350) \end{matrix}$
       & $\!\!\!\begin{matrix}   0.5241, \\ (28.28,1.0707) \end{matrix}$ & $\!\!\!\begin{matrix} {0.5372}, \\ (1.120,1.0424,\\ 0.800) \end{matrix}$
       & $\begin{matrix} {0.7614}, \\ (9.34,0.60,\\0.11,0.5947) \end{matrix}$ & $\begin{matrix} 0.6394, \\ 
       (2.85,2.75,\\ 0.8421) \end{matrix}$
        & $\!\!\!\!\begin{matrix} {0.7113}, \\ (1.485\!\times \!10^{-19},\\11.820,0.1090,\\0.4561) \end{matrix}$\\ \hline

$\mathfrak{B}_{\rm AAL}$ & $0.9790$ & $\!\!\!\!\begin{matrix}  0.9348, \\ (-0.1635,\\ 0.1635,0.9232) \end{matrix}$
        & $\!\!\!\begin{matrix}  0.8428, \\ (0.63,1.5873,\\0.63) \end{matrix}$
       & $\begin{matrix}   0.7836, \\ (171.62,\\1.0117) \end{matrix}$ & $\!\!\!\begin{matrix} {0.8191}, \\ (1.225,1.0686,\\ 0.6734) \end{matrix}$
       & $\begin{matrix} {0.9440}, \\ (10.4,0.61,\\0.11,0.5361) \end{matrix}$ & $\begin{matrix} 0.8907, \\ 
       (3.10,3.25,\\ 0.7295) \end{matrix}$
        & $\!\!\!\!\begin{matrix} {0.9243}, \\ (2.248\!\times \!10^{-16},\\11.820,0.1268,\\0.7306) \end{matrix}$\\ \hline
        
$\mathfrak{B}_{\rm AAL}$ & $1.8482$ & $\!\!\!\!\begin{matrix}  0.9928, \\ (-0.3826,\\ 0.3826,3.6200) \end{matrix}$
        & $\!\!\!\begin{matrix}  0.9427, \\ (0.995,1.005,\\0.995) \end{matrix}$
       & $\!\!\begin{matrix}   0.9719, \\ (2.98,1.6711) \end{matrix}$ & $\!\!\!\begin{matrix} {0.9479}, \\ (0.893,0.9463,\\ 1.2590) \end{matrix}$
       & $\begin{matrix} {0.9829}, \\ (4.7,0.41,\\0.11,2.5957) \end{matrix}$ & $\!\!\begin{matrix} 0.9868, \\ 
       (1.85,2,1.3108) \end{matrix}$
        &$\!\!\!\begin{matrix} {0.9971}, \\ (1.407\!\times \!10^{-7},\\3.764,0.1942,\\1.5676) \end{matrix}$ \\ \hline

$\mathfrak{B}_{\rm AAL}$ & $2.3285$ & $\!\!\!\!\begin{matrix}  0.9988, \\ (-0.2979,\\ 0.2979,2.2923) \end{matrix}$
        & $\!\!\!\begin{matrix}  0.9614, \\ (0.72,1.3889,\\0.72) \end{matrix}$
       & $\!\!\begin{matrix}   0.9586, \\ (5.54, 1.361) \end{matrix}$ & $\!\!\!\begin{matrix} {0.9503}, \\ (1.081,1.0303,\\ 0.8572) \end{matrix}$
       & $\begin{matrix} {0.9920}, \\ (2.10,2.11,\\0.71,0.1234) \end{matrix}$ & $\begin{matrix} 0.9892, \\ 
       (2.35,2.25,\\ 1.0591) \end{matrix}$
        & $\!\!\!\!\begin{matrix} {0.9984}, \\ (2.199\!\times \!10^{-28},\\10.92,0.0782,\\1.5885) \end{matrix}$ \\ \hline
        
$\mathfrak{B}^{\rm BC}_{\rm AAL}$ & $0.0047$ & $\!\!\!\!\begin{matrix}  0.7985, \\ (-0.0013,\\ 0.0013,0.0052) \end{matrix}$
        & $\begin{matrix}  0.8169, \\ (0.0052,\\192.3077,\\0.0052) \end{matrix}$
       & ${\rm Not~Valid}$ & $\!\!\!\!\!\begin{matrix} {0.9693}, \\ (17.059,\!1.0316,\\ 0.0052) \end{matrix}$
       & $\!\!\!\begin{matrix} {0.9505}, \\ (1.1472,\\15.3838,\\1.0196,0.0047) \end{matrix}$ & ${\rm Not Valid}$
        &$\!\!\!\!\begin{matrix} {0.9761}, \\ (\!1.057,\!15.0337,\\22.5118,\\0.0139) \end{matrix}$ \\ \hline

$\mathfrak{B}^{\rm BC}_{\rm AAL}$ & $0.0144$ & $\!\!\!\!\begin{matrix}  0.7399, \\ (-0.0036,\\ 0.0036,0.0145) \end{matrix}$
        & $\begin{matrix}  0.7753, \\ (0.0144,\\69.4444,\\0.0144) \end{matrix}$
       & ${\rm Not~Valid}$ & $\!\!\!\begin{matrix} {0.9632}, \\ (10.24,1.0501,\\ 0.0138) \end{matrix}$
       & $\begin{matrix} {0.9022}, \\ ( 6.755,4.535,\\0.860,0.0108) \end{matrix}$ & $\begin{matrix} 0.6683, \\ 
       (79.16,79.14,\\ 0.0254) \end{matrix}$
        &$\!\!\!\!\begin{matrix} {0.9921}, \\ (\!1.1223,\!8.1114,\\17.9978,\\0.0174) \end{matrix}$ \\ \hline
        
$\mathfrak{B}^{\rm BC}_{\rm AAL}$ & $0.0628$ & $\!\!\!\!\begin{matrix}  0.9929, \\ (-0.0161,\\ 0.0161,0.0665) \end{matrix}$
        & $\begin{matrix}   0.9901
        , \\ (0.0641,\\15.6006,\\0.0641) \end{matrix}$
       & ${\rm Not~Valid}$ & $\!\!\!\begin{matrix} {
           0.8857}, \\ (4.33, 1.0983,\\ 0.0682) \end{matrix}$
       & $\begin{matrix} {0.9859}, \\ ( 9.08,1.68,\\0.54,0.0651) \end{matrix}$ & $\!\!\!\begin{matrix} 0.9936, \\ 
       (31,31,0.0656) \end{matrix}$
        &$\!\!\!\!\begin{matrix} {0.9933}, \\ (\!0.0018,\!25.820,\\0.5970,\\0.0659) \end{matrix}$ \\ \hline

$\mathfrak{B}^{\rm BC}_{\rm AAL}$ & $0.0986
$ & $\!\!\!\!\begin{matrix}  0.7676, \\ (-0.0260,\\ 0.0260,0.1096) \end{matrix}$
        & $\!\!\!\begin{matrix}   0.8696
        , \\ \!(0.1050,\!9.5238,\\0.1050) \end{matrix}$
       & ${\rm Not~Valid}$ & $\!\!\!\begin{matrix} {
           0.9844}, \\ (3.60,1.1097,\\ 0.0952) \end{matrix}$
       & $\begin{matrix} {0.9144}, \\ ( 9.130,1.325,\\0.505,0.1057) \end{matrix}$ & $\begin{matrix}  0.8605, \\ 
       (167.15,9.95,\\ 0.1071) \end{matrix}$
        &$\!\!\!\!\begin{matrix} {0.9913}, \\ (\!1.2432,\!3.1535,\\4.6772,\!0.0959) \end{matrix}$  \\ \hline
        
$\mathfrak{B}^{\rm BC}_{\rm AAL}$ & $0.2082$ & $\!\!\!\!\begin{matrix}  0.9453, \\ (-0.0611,\\ 0.0611,0.2769) \end{matrix}$
        & $\!\!\!\begin{matrix}   0.9877
        , \\ \!(0.2350,\!4.2553,\\0.2350) \end{matrix}$
       & ${\rm Not~Valid}$ & $\!\!\!\begin{matrix} {
           0.9533}, \\ (2.27,1.1289,\\ 0.2177) \end{matrix}$
       & $\begin{matrix} {0.9847}, \\ ( 2.05,1.55,\\0.80,0.2278) \end{matrix}$ & $\!\!\!\begin{matrix}  0.9852, \\ 
       (35,4.8,0.2429) \end{matrix}$
        &$\!\!\!\!\begin{matrix} {0.9878}, \\ (\!0.2773,\!4.0760,\\1.0516,\!0.2160) \end{matrix}$  \\ \hline

$\mathfrak{B}^{\rm BC}_{\rm AAL}$ & $0.4786$ & $\!\!\!\!\begin{matrix}  0.9979, \\ (-0.0958,\\ 0.0958,0.4670) \end{matrix}$
        & $\!\!\!\begin{matrix}   0.9649
        , \\ \!(0.3174,\!3.1506,\\0.3174) \end{matrix}$
       & $\!\!\!\begin{matrix}   0.5170
               , \\ \!(\!171.62,\!1.0117) \end{matrix}$ & $\!\!\!\begin{matrix} {
           0.9082}, \\ (1.908, 1.1271,\\  0.2974) \end{matrix}$
       & $\begin{matrix} {0.9868}, \\ ( 2.25,1.35,\\0.65,0.2762) \end{matrix}$ & $\begin{matrix}  0.9877, \\ 
       ( 5.90,5.85,\\0.3694) \end{matrix}$
        & $\!\!\!\!\begin{matrix} {0.9956}, \\ (1.4649\!\times\!10^{-9}\!,\\12.9900,\\0.2054,\!0.4262) \end{matrix}$ \\ \hline
\hline
 		\end{tabular}
 		\vspace{-0.2in}
 \end{table*}
 
 Fig. \ref{fig_two}(a) illustrates the received signal through a fresh water link without any artificially-induced turbulence. As it can be seen, for a UWOC channel with low link range (like our water tank) the received signal is approximately constant over a large period of time, i.e., fading has a negligible effect on the performance of low-range UWOC systems. As a consequence, the channel impulse response for such scenarios can be thoroughly described by a deterministic FFIR. Moreover, the corresponding histogram of the acquired data in Fig. \ref{fig_three}(a) and the simulation results in the first row of Table I show that for such a channel condition the normalized received optical power is mainly confined around the mean of ``$1$" with a negligible standard deviation and hence lots of the discussed statistical distributions in Sec. IV give an excellent match to the data histogram\footnote{Note that the parameter $\alpha$ in the K distribution is positive and hence, based on the relation $\sigma^2_I=1+2/\alpha$,  the evaluation of K distribution is only valid for the channels with $\sigma^2_I>1$. Moreover, for very small values of $\sigma^2_I$, the parameter $\beta$ in the exponentiated Weibull and the parameters $\alpha$ and $\beta$ in the Gamma-Gamma distribution have large values hampering the analytical evaluation of exponentiated Weibull and Gamma-Gamma distributions for such regimes of the scintillation index.}.
 
 According to Fig. \ref{fig_two}(a), due to the low link range available by the laboratory water tank, the received signal fluctuations are typically negligible. In other words, optical turbulence manifests its effect at longer link ranges, i.e., for UWOC channels with typically $d_0>10$ \si{m} link ranges \cite{korotkova2012light}; that is why we used some procedures to artificially induce fluctuations on the propagation path.
 Figs. \ref{fig_two}(b)-(e) illustrate the received signal through a bubbly fresh water link with different concentrations of air bubbles, i.e., different fading strengths, when neither the transmitter BC nor the receiver AAL is employed in the link configuration (a $\mathfrak{B}$ link).
 As it is observed, the presence of air bubbles within the channel causes the received signal to severely fluctuate. This is mainly due to the random arrangement of air bubbles through the propagation path causing the propagating photons to randomly scatter in different directions and leave their direct path.
 The corresponding histograms of the acquired experimental data and also the accordance of different statistical distributions are depicted in Figs. \ref{fig_three}(b)-(e). From these figures and the detailed numerical results in the second-fifth rows of Table I, we can observe that while a bubbly UWOC link under a relatively strong fading condition, like $\sigma^2_I=0.7606$ and $3.3415$, can be well described by many of the well-known statistical distributions (approximately all of the seven statistical distributions considered in this paper), for moderate fading conditions, like $\sigma^2_I=0.2408$, non of the statistical distributions under consideration can acceptably model the fading behavior.
 The major reason can intuitively be deduced from the temporal representation of the received optical signal in Figs. \ref{fig_two}(b)-(e). Based on these figures, the presence of air bubbles causes the received intensity to mainly lie either in large or small values. This is mainly because in the current link configuration neither the transmitter BC nor the receiver AAL is used and hence the received intensity is very sensitive to the beam scattering caused by the random presence of air bubbles. Therefore, as we showed in our previous experimental research \cite{jamali2016statistical}, in such circumstances the typical single-lobe distributions cannot appropriately fit the experimental data and generally a two-lobe statistical distribution, such as the mixed exponential-lognormal distribution we have proposed in \cite{jamali2016statistical}, is required to predict the statistical behavior of UWOC fading in all regions of the scintillation index.
 
 In order to make the link geometry commensurate with the practical scenarios and in the meantime reduce the above-discussed sensitivity of the received optical signal to the beam scattering, hereafter, for all of the considered scenarios, we assume that the receiver is equipped with an AAL. Figs. \ref{fig_two}(f)-(i) show the received signal through a bubbly fresh water link with different concentrations of air bubbles when just the receiver AAL is employed, i.e., a $\mathfrak{B}_{\rm AAL}$ link. From the corresponding histograms of the acquired experimental data depicted in Figs. \ref{fig_three}(f)-(i), one can observe that the application of AAL at the receiver side not only effectively reduces the fading strength measured by the scintillation index but also makes the fluctuations on the received optical signals through the turbulent UWOC channels predictable by the well-known single-lobe statistical distributions.
 In particular, based on the detailed numerical results in the sixth-ninth rows of Table I, we can observe that relatively all the seven statistical distributions can acceptably model the fading behavior; however, the generalized Gamma and exponentiated Weibull distributions, respectively, always give the best accordance to the acquired experimental data in terms of their GoF.
 
 {Before experimentally exploring the transmitter BC effect on the UWOC fading behavior, it is worth to first briefly clarify the functionalities of the  transmitter BC and receiver AAL from the theoretical point of view. In fact, while an optical receiver with a very small aperture will result in a random signal, increasing the receiver aperture lens to larger than that of the spatial scale of the irradiance fluctuations, significantly reduces the signal fluctuations compared to a point receiver through averaging the fluctuations over the aperture which is well known as aperture averaging effect in the literature \cite{churnside1991aperture}. On the other hand, the expanded beam by the transmitter BC can be thought as of a collection of spatially separated rays each experiencing somewhat  independent fluctuations when propagating through the random UWOC channel. Systematically speaking, AAL and BC attempt to somehow realize the receiver and transmitter diversity, respectively. Therefore, these two components together somehow perform in a similar fashion to MIMO transmission which significantly reduces the optical turbulence through spatial diversity by providing so-called independent paths from the transmitter to the receiver. Accordingly, we expect that these two components significantly reduce the turbulence strength measured by the scintillation index through the basic fundamentals of spatial diversity in optical turbulent channels \cite{jamali2015performanceMIMO}. In addition, since BC and AAL significantly reduce the deep fade probability (or equivalently, considerably reduce the link sensitivity to beam wandering), the quantized measured samples will no longer only accumulate either at the large or small values; instead, they will mainly tend to take the values around the mean value (normalized by one).
 	
 We should emphasize that because of the negligibility of background noise in UWOC systems deployed in typical water depths \cite{jamali2015ber}, the application of AAL sounds to be useful in most of the cases since it not only mitigates the turbulence impairments but also collects lots of the scattered photons onto its focal plane where the photo-detector is placed. On the other hand, BC usage at the transmitter side sometimes comes with the cost of reduced power reached to the receiver. In fact, when we expand the cross-section of the sharp optical beam of the laser using the transmitter BC, we expect the number of interactions between the propagating photons and water suspended particles and molecules significantly increase; this results in a remarkably higher number of absorption and scattering events and increases the total loss on the received optical power. However, 
 depending on the channel condition, including the water type and turbulence strength, the aforementioned cost may be negligible. For example, in the turbid harbor water links the multiple-scattering effect of the channel is too much such that, even for the relatively short link range of $10$ m, the sharp beam of transmitter laser will be spread over a wide spatial area at the receiver plane; hence, the received pattern in both temporal and spatial domains (including the attenuation, delay spread, and spatial dispersion) tends to be similar to that of LED-based diffusive links \cite{jamali2017mimovlc}. This implies that for such link conditions, using either sharp laser or diffusive LED sources results in a similar received pattern over a real UWOC channel with a typical length, which itself signifies that using the transmitter BC will negligibly change the channel attenuation. Similar inferences can be derived for any other channel conditions.}
 
Now, in order to experimentally investigate the effect of beam-collimator (and expander) on the statistics of the received optical signal through turbulent UWOC channels, as described in Sec. V, we also run a comprehensive set of experiments when the link geometry possesses such an integral part. Figs. \ref{fig_two}(j)-(o) show the received signal through a bubbly fresh water link with different concentrations of air bubbles when both the transmitter BC and the receiver AAL are employed, i.e., a $\mathfrak{B}^{\rm BC}_{\rm AAL}$ link. The corresponding histograms of the acquired experimental data depicted in Figs. \ref{fig_three}(j)-(o) demonstrate that employing the transmitter BC in conjunction with the receiver AAL significantly reduces the fading strength such that in the highest amount of bubble concentration the scintillation index is considerably less than one. On the other hand, such a link configuration also remarkably causes the received intensity samples to concentrate around the mean of the acquired data, i.e., a relatively apposite behavior to $\mathfrak{B}$ links is observable such that the acquired data samples, with a good probability, are no longer only either at large or small values. As a consequence, and based on the detailed numerical results in the tenth-fifteenth rows of Table I, the fading in $\mathfrak{B}^{\rm BC}_{\rm AAL}$ links can appropriately be characterized using simple single-lobe statistical distributions; however, once again, the generalized Gamma and exponentiated Weibull distributions, respectively, yield much better accordance to the acquired experimental data in terms of their GoF.

 \subsection{Turbulence-Induced Fading due to the Temperature Random Variations}
 \begin{figure*}
   \vspace{-0.1in}
      \centering
      \includegraphics[width=7in,height=7in]{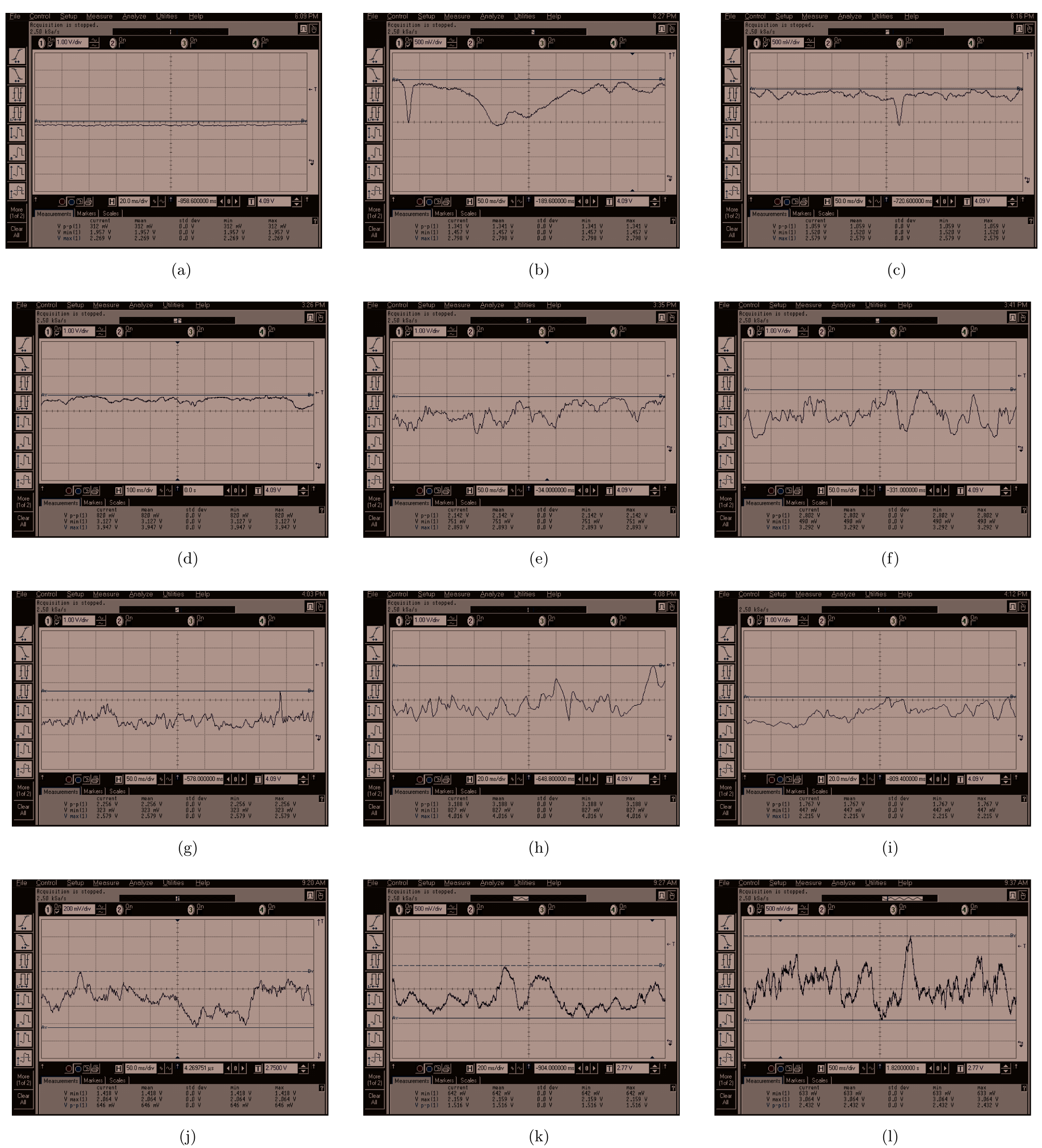}
      \caption{Received optical signal through (a) a UWOC link, under the temperature random variations caused by circular heating elements, equipped with both the transmitter BC and the receiver AAL, $\mathcal{H}^{\rm BC}_{\rm AAL}$, with $\sigma^2_I=9.2395 \times 10^{-4}$; (b) a $\mathcal{H}_{\rm AAL}$ link with $\sigma^2_I=0.0046$;  (c) a $\mathcal{H}_{\rm AAL}$ link with $\sigma^2_I=0.0072$; (d) a UWOC link, under the temperature random variations caused by tunable water droppers, equipped with just the receiver AAL, $\mathfrak{D}_{\rm AAL}$, with $\sigma^2_I=0.0021$; (e) a $\mathfrak{D}_{\rm AAL}$ link with $\sigma^2_I=0.0577$; (f) a $\mathfrak{D}_{\rm AAL}$ link with $\sigma^2_I=0.1108$; (g) a $\mathfrak{D}_{\rm AAL}$ link with $\sigma^2_I=0.1647$; (h) a $\mathfrak{D}_{\rm AAL}$ link with $\sigma^2_I=0.2611$; (i) a $\mathfrak{D}_{\rm AAL}$ link with $\sigma^2_I=0.3668$; (j) a $\mathfrak{D}^{\rm BC}_{\rm AAL}$ link with $\sigma^2_I=0.0284$; (k) a $\mathfrak{D}^{\rm BC}_{\rm AAL}$ link with $\sigma^2_I=0.0632$; and (l) a $\mathfrak{D}^{\rm BC}_{\rm AAL}$ link with $\sigma^2_I=0.0897$.}
      \vspace{-0.2in}
      \label{fig_four}
      \end{figure*}
 \begin{figure*}
         \centering
         \includegraphics[trim=0cm 0cm 0cm 0.5cm,width=6.8in,height=6.6in,clip]{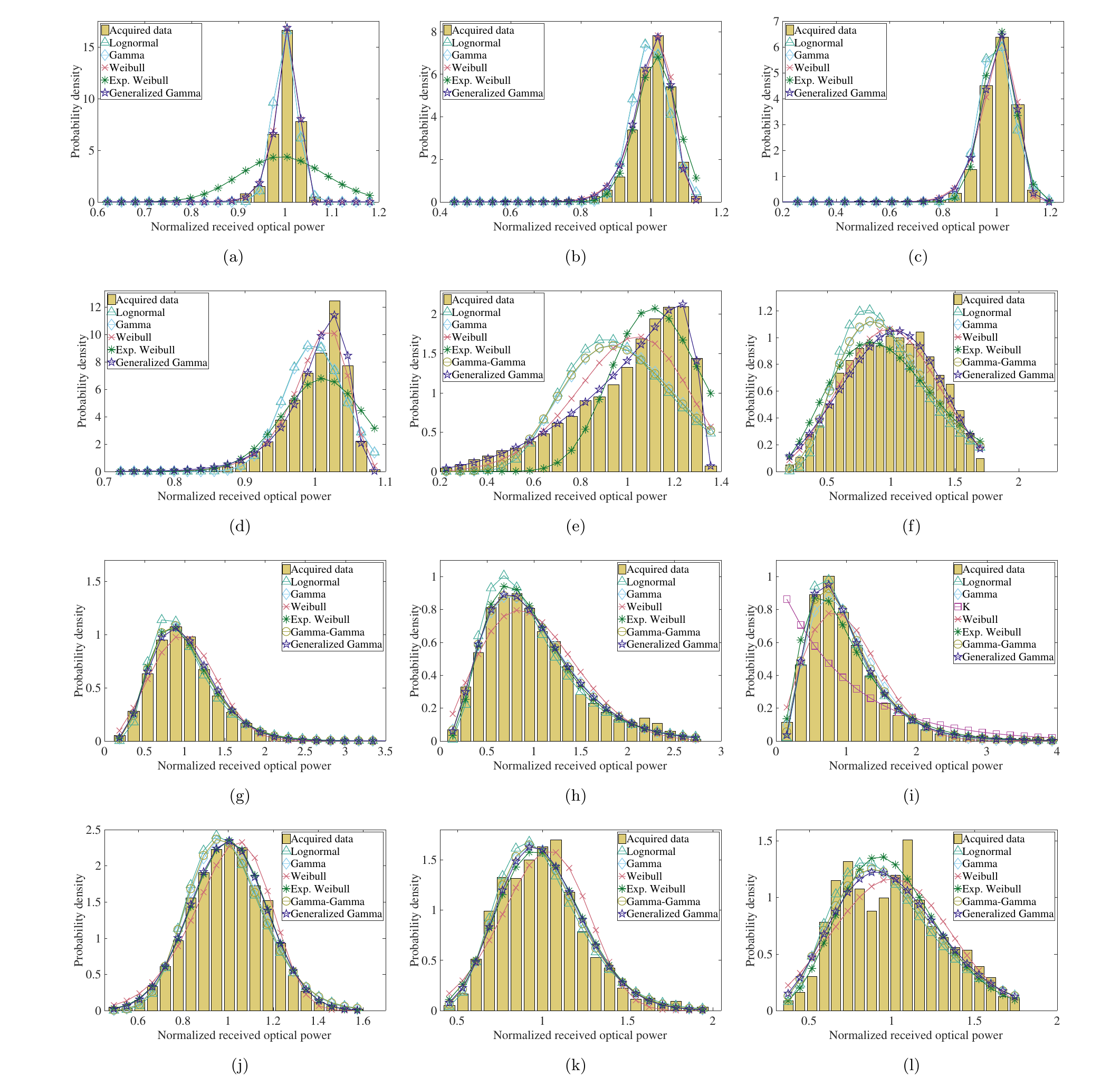}
         \caption{Accordance of different statistical distributions, considered in Sec. IV, with the histograms of the acquired data through various UWOC channel conditions under the temperature random variations, namely (a) a UWOC link, under the temperature random variations caused by circular heating elements, equipped with both the transmitter BC and the receiver AAL, $\mathcal{H}^{\rm BC}_{\rm AAL}$, with $\sigma^2_I=9.2395 \times 10^{-4}$; (b) a $\mathcal{H}_{\rm AAL}$ link with $\sigma^2_I=0.0046$;  (c) a $\mathcal{H}_{\rm AAL}$ link with $\sigma^2_I=0.0072$; (d) a UWOC link, under the temperature random variations caused by tunable water droppers, equipped with just the receiver AAL, $\mathfrak{D}_{\rm AAL}$, with $\sigma^2_I=0.0021$; (e) a $\mathfrak{D}_{\rm AAL}$ link with $\sigma^2_I=0.0577$; (f) a $\mathfrak{D}_{\rm AAL}$ link with $\sigma^2_I=0.1108$; (g) a $\mathfrak{D}_{\rm AAL}$ link with $\sigma^2_I=0.1647$; (h) a $\mathfrak{D}_{\rm AAL}$ link with $\sigma^2_I=0.2611$; (i) a $\mathfrak{D}_{\rm AAL}$ link with $\sigma^2_I=0.3668$; (j) a $\mathfrak{D}^{\rm BC}_{\rm AAL}$ link with $\sigma^2_I=0.0284$; (k) a $\mathfrak{D}^{\rm BC}_{\rm AAL}$ link with $\sigma^2_I=0.0632$; and (l) a $\mathfrak{D}^{\rm BC}_{\rm AAL}$ link with $\sigma^2_I=0.0897$.}
         \vspace{-0.15in}
         \label{fig_five}
         \end{figure*}
     \begin{table*}
     		\centering
     		\caption{GoF and the constant parameters of different PDFs for the various channel conditions considered in Scenario II (Sec. VI-B).}
     		\begin{tabular}{||p{0.45in}||p{0.35in}||p{0.6in}||p{0.6in}||p{0.6in}||p{0.6in}||p{0.6in}||p{0.7in}||p{0.6in}||}
     			\hline\hline
     			\! Channel condition \vspace{-2cm} & \vspace{0.25cm} $\sigma^2_{I,{m}}$& 
    \vspace{0.1cm} {Log-normal}&
    \vspace{0.1cm} {Gamma}&
    \vspace{0.1cm} {K dist.}&
    \vspace{0.1cm} {Weibull}&
    \vspace{0.1cm} $\!\!${Exp. Weibull}&
    \vspace{0.1cm} $\!\!${Gamma-Gamma}&
    \vspace{0.1cm} $\!\!${Generalized Gamma}\\
     			\cline{3-9}
    && $\begin{matrix} \!{\rm GoF}, \\ \!(\mu_X,\sigma^2_X,\sigma^2_I) \end{matrix}$ &
     $\begin{matrix} {\rm GoF}, \\ (\theta,k,\sigma^2_I) \end{matrix}$ & 
     $\begin{matrix} {\rm GoF}, \\ (\alpha,\sigma^2_I) \end{matrix}$ & 
    $\begin{matrix} {\rm GoF}, \\ (\beta,\eta,\sigma^2_I) \end{matrix}$ &
    $\begin{matrix} {\rm GoF}, \\ (\alpha,\beta,\eta,\sigma^2_I) \end{matrix}$ & 
     $\begin{matrix} {\rm GoF}, \\ (\alpha,\beta,\sigma^2_I) \end{matrix}$ & 
    $\begin{matrix} {\rm GoF}, \\ (a,d,p,\sigma^2_I) \end{matrix}$ 
    \\ \hline \hline 			
     			$\mathcal{H}^{\rm BC}_{\rm AAL}$ &$9.2395 \times 10^{-4}$ & $\begin{matrix} \!{0.9607}, \\ \!\!\!\!(-1.4\!\times\!\!10^{-4},\\ 1.4\!\times\!\! 10^{-4},\\ \!\!5.6016\!\times\!\! 10^{-4}) \end{matrix}$
     			 & $\begin{matrix} \!{0.9629}, \\ \!\!\!\!(5.82\!\times\!\!10^{-4},\\ 1.718\!\times\!\! 10^{3},\\ 5.82\!\times\!\! 10^{-4}) \end{matrix}$
     			  & ${\rm Not~Valid}$ & $\begin{matrix} \!{0.9981}, \\ \!\!\!\!(48.663,\\1.0116,\\ 6.746\!\times\!\! 10^{-4}) \end{matrix}$
     			  & $\!\!\!\begin{matrix} 0.3692, \\ (7.2710,5,\\ 0.0084,\!0.0084) \end{matrix}$ & ${\rm Not~Valid}$ &$\!\!\!\!\begin{matrix} {0.9983}, \\ (\!1.0173,\!45.245,\\58.3382,\\\!0.0046) \end{matrix}$  \\ \hline

$\mathcal{H}_{\rm AAL}$&$0.0046$ & $\begin{matrix} {0.9413}, \\ \!\!\!\!(-6.9 \times 10^{-4},\\ 6.9\times 10^{-4},\\ 0.0028) \end{matrix}$
     			 & $\begin{matrix} 0.9475, \\ (0.0028,\\ 361.0108,\\0.0028) \end{matrix}$
     			  & ${\rm Not~Valid}$ & $\!\!\!\!\begin{matrix} 0.9922, \\ (21.803,\!1.0251,\\ 0.0032) \end{matrix}$
     			  & $\begin{matrix} 0.9704, \\ (7.30,7.94,\\ 0.91, 0.0033) \end{matrix}$ & ${\rm Not~Valid}$ &$\!\!\!\!\begin{matrix} {0.9949}, \\ (\!0.9991,\!25.202,\\17.3830,\\\!0.0031) \end{matrix}$   \\ \hline
     			  
$\mathcal{H}_{\rm AAL}$ &$0.0072$ & $\begin{matrix} {0.9576}, \\ \!\!\!\!(-9.6 \times 10^{-4},\\ 9.6\times 10^{-4},\\ 0.0038) \end{matrix}$
     			 & $\begin{matrix} 0.9633, \\ (0.0039,\\ 258.3979,\\0.0039) \end{matrix}$
     			  & ${\rm Not~Valid}$ & $\begin{matrix} 0.9919, \\ ( 18.4790,\\1.0293,\\ 0.0045) \end{matrix}$
     			  & $\begin{matrix} 0.9925, \\ (7.30,7.74,\\ 0.90, 0.0035) \end{matrix}$ & ${\rm Not~Valid}$ & $\!\!\!\!\begin{matrix} {0.9950}, \\ (\!0.9850,\!22.325,\\13.4807,\\\!0.0042) \end{matrix}$\\ \hline
     			  
$\mathfrak{D}_{\rm AAL}$ &$0.0021$ & $\begin{matrix} {0.8145}, \\ \!\!\!\!(-4.6 \times 10^{-4},\\ 4.6\times 10^{-4},\\ 0.0018) \end{matrix}$
     			 & $\begin{matrix} 0.8227, \\ (0.0018,\\ 546.4481,\\0.0018) \end{matrix}$
     			  & ${\rm Not~Valid}$ & $\begin{matrix} 0.9620, \\ ( 28.9090,\\1.0192,\\  0.0019) \end{matrix}$
     			  & $\!\!\!\begin{matrix} 0.7747, \\ (7.300,7.865,\\ 0.905, 0.0034) \end{matrix}$ & ${\rm Not~Valid}$ & $\!\!\!\!\begin{matrix} {0.9851}, \\ (\!1.0438,\!22.993,\\53.0739,\\\!0.0022) \end{matrix}$ \\ \hline
     			  
$\mathfrak{D}_{\rm AAL}$ &$0.0577$ & $\!\!\!\begin{matrix} {0.3321}, \\ \!\!\!\!(-0.0159,\\ 0.0159,\!0.0658) \end{matrix}$
     			 & $\!\!\!\begin{matrix} 0.4394, \\ \!(0.0647,\!15.456,\\0.0647) \end{matrix}$
     			  & ${\rm Not~Valid}$ & $\!\!\!\begin{matrix} 0.7137, \\ ( 4.970,1.0895,\\0.0530) \end{matrix}$
     			  & $\!\begin{matrix} 0.7276, \\ (7.02,2.70,\\ 0.82, 0.0296) \end{matrix}$ & $\!\begin{matrix} 0.4214, \\ (157.02,16.62,\\ 0.0669) \end{matrix}$ & $\!\!\!\!\begin{matrix} {0.9941}, \\ (\!1.3107,\!3.3696,\\40.129,\!0.0568) \end{matrix}$  \\ \hline

$\mathfrak{D}_{\rm AAL}$ &$0.1108$ & $\!\!\!\begin{matrix} {0.6351}, \\ \!\!\!\!(-0.0365,\\ 0.0365,\!0.1572) \end{matrix}$
     			 & $\!\!\!\begin{matrix} 0.8054, \\ \!(0.145,\!6.8966,\\0.145) \end{matrix}$
     			  & ${\rm Not~Valid}$ & $\!\!\!\begin{matrix} 0.9452, \\ \!( 3.0426,\!1.1191,\\0.1288) \end{matrix}$
     			  & $\!\begin{matrix} 0.7810, \\ (2.00,1.74,\\0.84,0.1861) \end{matrix}$ & $\!\begin{matrix} 0.7948, \\ (170.25,7.1500,\\0.1466) \end{matrix}$ & $\!\!\!\!\begin{matrix} {0.9570}, \\ (\!1.3049,\!2.6108,\\4.173,\!0.1325) \end{matrix}$\\ \hline
     			  
$\mathfrak{D}_{\rm AAL}$ &$0.1647$ & $\!\!\!\begin{matrix} {0.9709}, \\ \!\!\!\!(-0.0397,\\ 0.0397,\!0.1721) \end{matrix}$
     			 & $\!\!\!\begin{matrix} 0.9956, \\ \!(0.156,\!6.4103,\\0.156) \end{matrix}$
     			  & ${\rm Not~Valid}$ & $\!\!\!\begin{matrix} 0.973, \\ \!( 2.8103,\!1.1229,\\0.1485) \end{matrix}$
     			  & $\!\begin{matrix} 0.9932, \\ (4.319,1.35,\\0.582,0.1647) \end{matrix}$ & $\begin{matrix} 0.9949, \\ (165.1,6.7,\\0.1562) \end{matrix}$ & $\!\!\!\!\begin{matrix} {0.9968}, \\ (\!0.3557,\!5.0965,\\1.296,\!0.1522) \end{matrix}$\\ \hline
     			  
$\mathfrak{D}_{\rm AAL}$ &$0.2611$ & $\!\!\!\begin{matrix} {0.9567}, \\ \!\!\!\!(-0.0675,\\ 0.0675,\!0.3100) \end{matrix}$
     			 & $\!\!\!\begin{matrix} 0.9883, \\ \!(0.260,3.8462,\\0.260) \end{matrix}$
     			  & ${\rm Not~Valid}$ & $\!\!\!\begin{matrix} 0.9422, \\ \!( 2.1393,\!1.1292,\\0.2421) \end{matrix}$
     			  & $\!\!\!\!\begin{matrix} 0.9845, \\ (4.8246,\!1.0395,\\0.4604,\!0.2611) \end{matrix}$ & $\begin{matrix} 0.9891, \\ (74.5,4.0,\\0.2668) \end{matrix}$ &$\!\!\!\!\begin{matrix} {0.9890}, \\ (\!0.1503,\!4.3724,\\0.863,\!0.2431) \end{matrix}$ \\ \hline
     			  
$\mathfrak{D}_{\rm AAL}$ &$0.3668$ & $\!\!\!\begin{matrix} {0.9916}, \\ \!\!\!\!(-0.0679,\\ 0.0679,\!0.3121) \end{matrix}$
     			 & $\!\!\!\begin{matrix} 0.9746, \\ \!(0.262,3.8168,\\0.262) \end{matrix}$
     			  & $\begin{matrix} 0.3649, \\(171.62,\\1.0117) \end{matrix}$
     			   & $\!\!\begin{matrix} 0.9148, \\ \!( 2.0964,\!1.129,\\0.2511) \end{matrix}$
     			  & $\!\!\!\!\begin{matrix} 0.9692, \\ (5.1679,\!0.8677,\\0.3721,\!0.3668) \end{matrix}$ & $\!\!\!\begin{matrix} 0.9889, \\ (7.5,7.5,0.2844) \end{matrix}$ &$\!\!\!\begin{matrix} {0.9916}, \\ (3.142\!\times\!10^{-7},\\13.6600,\\0.265,\!0.2843) \end{matrix}$ \\ \hline

$\mathfrak{D}^{\rm BC}_{\rm AAL}$ &$ 0.0284$ & $\!\!\!\begin{matrix} {0.9685}, \\ \!\!\!\!(-0.0072,\\ 0.0072,\!0.0292) \end{matrix}$
     			 & $\!\!\!\begin{matrix} 0.984, \\ \!(0.029,34.482,\\ 0.0290) \end{matrix}$
     			  & ${\rm Not~Valid}$
     			   & $\!\!\begin{matrix} 0.9676, \\ \!( 6.717,\!1.0713,\\0.0305) \end{matrix}$
     			  & $\!\!\begin{matrix} 0.9956, \\ (2.440,\! 4.198,\\0.925,\!0.0284) \end{matrix}$ & $\!\!\!\begin{matrix} 0.9780, \\ (106.42,50.50,\\0.0294) \end{matrix}$ & $\!\!\!\!\begin{matrix} {0.9961}, \\ (0.6549,\!12.128,\\3.055,\!0.0289) \end{matrix}$\\ \hline
     			  
$\mathfrak{D}^{\rm BC}_{\rm AAL}$ & $0.0632$ & $\!\!\!\begin{matrix} {0.9585}, \\ \!\!\!\!(-0.0160,\\ 0.0160,\!0.0661) \end{matrix}$
     			 & $\!\!\begin{matrix} 0.9703, \\ \!(0.062,16.129,\\ 0.062) \end{matrix}$
     			  & ${\rm Not~Valid}$
     			   & $\!\!\!\begin{matrix} 0.9193, \\ \!(\! 4.6225,\!1.0941,\\0.0605) \end{matrix}$
     			  & $\!\!\begin{matrix} 0.9686, \\ (3.247,\! 2.447,\\0.804,\!0.0632) \end{matrix}$ & $\begin{matrix} 0.9691, \\ (148.2,17.9,\\0.0630) \end{matrix}$ & $\!\!\!\!\begin{matrix} {0.9713}, \\ (0.1945,\!12.238,\\1.3418,\!0.0620) \end{matrix}$\\ \hline
     			  
$\mathfrak{D}^{\rm BC}_{\rm AAL}$ & $0.0897$ & $\!\!\!\begin{matrix} {0.7969}, \\ \!\!\!\!(-0.0291,\\ 0.0291,\!0.1234) \end{matrix}$
     			 & $\!\!\begin{matrix} 0.8271, \\ \!(0.113,8.8496,\\ 0.113) \end{matrix}$
     			  & ${\rm Not~Valid}$
     			   & $\!\!\!\begin{matrix} 0.7906, \\ \!(\! 3.3579,\!1.1138,\\0.1079) \end{matrix}$
     			  & $\!\!\begin{matrix} 0.7916, \\ (3.633,\! 1.954,\\0.731,\!0.0897) \end{matrix}$ & $\begin{matrix} 0.8253, \\ (169.38,9.30,\\0.1141) \end{matrix}$ &$\!\!\!\!\begin{matrix} {0.8298}, \\ (0.3605,\!6.4043,\\1.434,\!0.1127) \end{matrix}$ \\ \hline
\hline
 		\end{tabular}
 \end{table*}
    \begin{figure*}
       \centering
       \includegraphics[trim=0cm 0cm 0cm 0.5cm,width=7in,height=3.2in,clip]{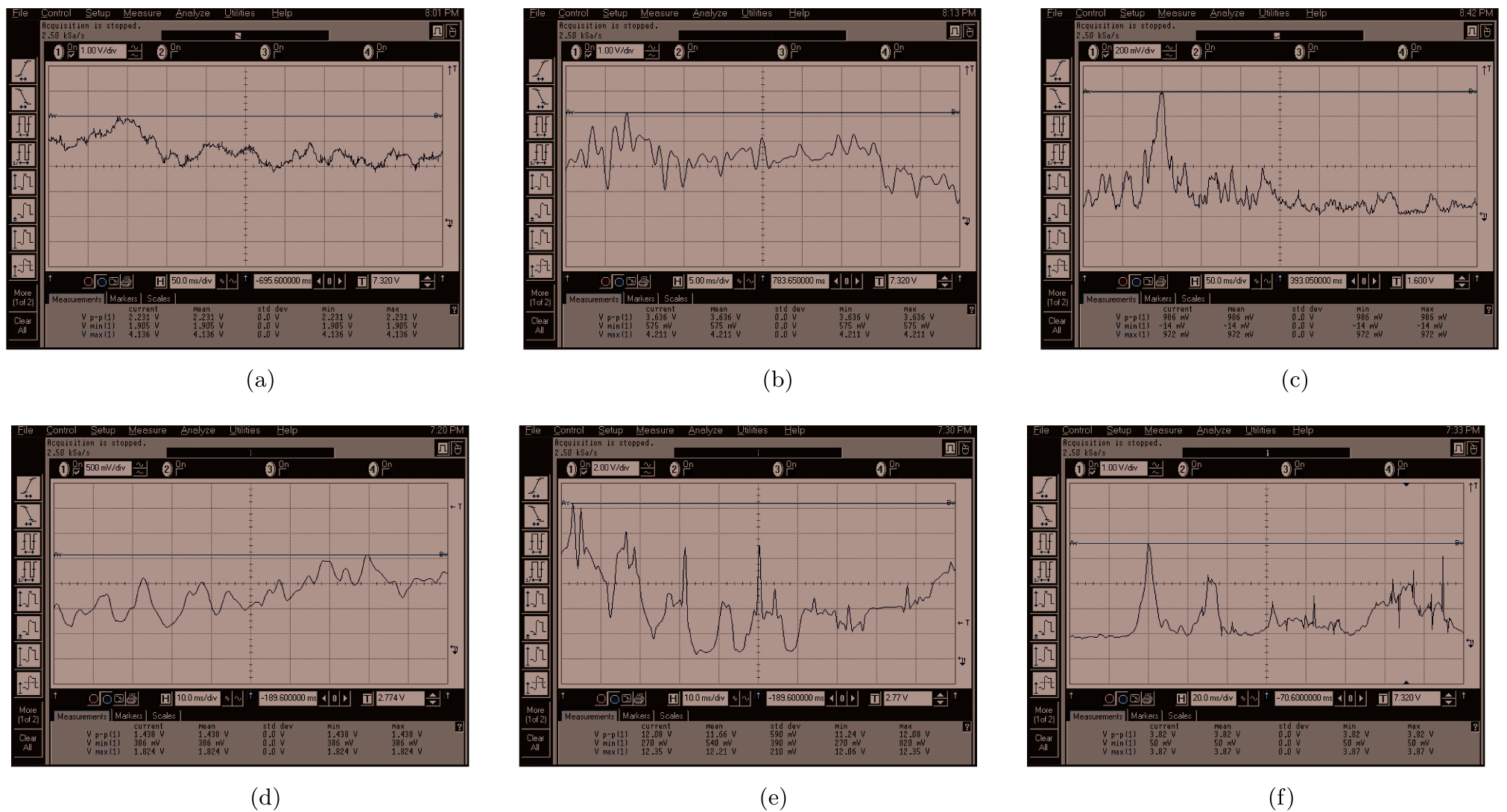}
       \caption{Received optical signal through (a) a UWOC link, under the mixture of temperature and air bubble random variations, equipped with both the transmitter BC and the receiver AAL, $\mathcal{M}^{\rm BC}_{\rm AAL}$, with $\sigma^2_I=0.0792$; (b) a $\mathcal{M}^{\rm BC}_{\rm AAL}$ link with $\sigma^2_I=0.3191$; (c) a $\mathcal{M}^{\rm BC}_{\rm AAL}$ link with $\sigma^2_I=0.7648$; (d) a $\mathcal{M}_{\rm AAL}$ link with $\sigma^2_I=0.1152$; (e) a $\mathcal{M}_{\rm AAL}$ link with $\sigma^2_I=0.4339$; and (f) a $\mathcal{M}_{\rm AAL}$ link with $\sigma^2_I=0.8862$.}
       \label{fig_six}
       \end{figure*}
       
       \begin{figure*}
              \centering
              \includegraphics[trim=0cm 0cm 0cm .2cm,width=7in,height=3.1in,clip]{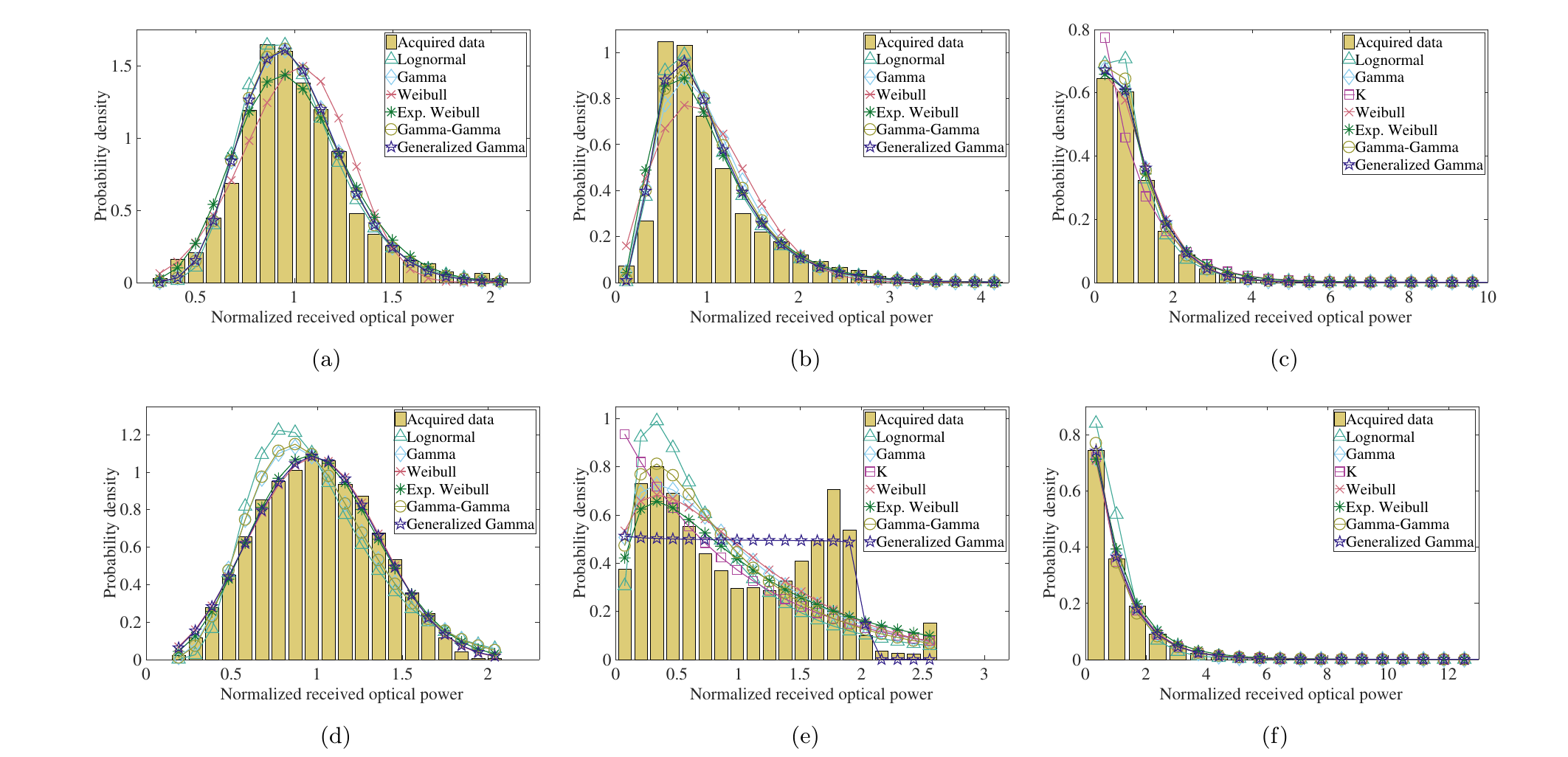}
              \caption{Accordance of different statistical distributions with the histograms of the acquired data through various UWOC channel conditions under the mixture of temperature and air bubble random variations, namely (a) a UWOC link, under the mixture of temperature and air bubble random variations, equipped with both the transmitter BC and the receiver AAL, $\mathcal{M}^{\rm BC}_{\rm AAL}$, with $\sigma^2_I=0.0792$; (b) a $\mathcal{M}^{\rm BC}_{\rm AAL}$ link with $\sigma^2_I=0.3191$; (c) a $\mathcal{M}^{\rm BC}_{\rm AAL}$ link with $\sigma^2_I=0.7648$; (d) a $\mathcal{M}_{\rm AAL}$ link with $\sigma^2_I=0.1152$; (e) a $\mathcal{M}_{\rm AAL}$ link with $\sigma^2_I=0.4339$; and (f) a $\mathcal{M}_{\rm AAL}$ link with $\sigma^2_I=0.8862$.}
              \vspace{-0.15in}
              \label{fig_seven}
              \end{figure*}
       
\begin{table*}
     		\centering
     		\caption{GoF and the constant parameters of different PDFs for the various mixtures of temperature and air bubble random variations.}
     		\begin{tabular}{||p{0.45in}||p{0.35in}||p{0.6in}||p{0.6in}||p{0.6in}||p{0.6in}||p{0.6in}||p{0.7in}||p{0.6in}||}
     			\hline\hline
     			\! Channel condition \vspace{-2cm} & \vspace{0.25cm} $\sigma^2_{I,{m}}$& 
    \vspace{0.1cm} {Log-normal}&
    \vspace{0.1cm} {Gamma}&
    \vspace{0.1cm} {K dist.}&
    \vspace{0.1cm} {Weibull}&
    \vspace{0.1cm} $\!\!${Exp. Weibull}&
    \vspace{0.1cm} $\!\!${Gamma-Gamma}&
    \vspace{0.1cm} $\!\!${Generalized Gamma}\\
     			\cline{3-9}
    && $\begin{matrix} \!{\rm GoF}, \\ \!(\mu_X,\sigma^2_X,\sigma^2_I) \end{matrix}$ &
     $\begin{matrix} {\rm GoF}, \\ (\theta,k,\sigma^2_I) \end{matrix}$ & 
     $\begin{matrix} {\rm GoF}, \\ (\alpha,\sigma^2_I) \end{matrix}$ & 
    $\begin{matrix} {\rm GoF}, \\ (\beta,\eta,\sigma^2_I) \end{matrix}$ &
    $\begin{matrix} {\rm GoF}, \\ (\alpha,\beta,\eta,\sigma^2_I) \end{matrix}$ & 
     $\begin{matrix} {\rm GoF}, \\ (\alpha,\beta,\sigma^2_I) \end{matrix}$ & 
    $\begin{matrix} {\rm GoF}, \\ (a,d,p,\sigma^2_I) \end{matrix}$ 
    \\ \hline \hline 			
     			$\mathcal{M}^{\rm BC}_{\rm AAL}$ &$0.0792$ & $\!\!\!\begin{matrix} 0.9774, \\ (-0.0162,\\0.0162,\!0.0669) \end{matrix}$
     			 & $\!\!\!\!\begin{matrix} 0.9836, \\ (0.065,\!15.3846\\0.065) \end{matrix}$
     			  & ${\rm Not~Valid}$ & $\!\!\!\!\begin{matrix} 0.9177, \\ (4.3395,\!1.0981,\\0.0679) \end{matrix}$
     			  & $\!\!\!\!\begin{matrix} 0.9668, \\ (3.494,2.116,\\0.7586,0.0792) \end{matrix}$ & $\begin{matrix} 0.9836, \\ (148.6,17.2,\\0.0653) \end{matrix}$ & $\!\!\!\!\begin{matrix} {0.9836}, \\ (0.0436,\!16.636,\\0.924,\!0.0649) \end{matrix}$  \\ \hline

$\mathcal{M}^{\rm BC}_{\rm AAL}$ &$0.3191$ & $\!\!\!\begin{matrix} 0.9751, \\ (-0.0658,\\0.0658,\!0.3011) \end{matrix}$
     			 & $\!\!\begin{matrix} 0.9206, \\ (0.255,\!3.9216\\0.255) \end{matrix}$
     			  & ${\rm Not~Valid}$ & $\!\!\!\!\begin{matrix} 0.8413, \\ (2.0646,\!1.1289,\\0.2581) \end{matrix}$
     			  & $\!\!\!\!\begin{matrix}  0.9424, \\ (5.031,0.933,\\0.4077,0.3191) \end{matrix}$ & $\!\!\!\begin{matrix} 0.9508, \\ (7.8,7.8,0.2728) \end{matrix}$ & $\!\!\!\begin{matrix} {0.9633}, \\ (1.773\!\times\!10^{-9},\\14.3200,\\0.259,\!0.2801) \end{matrix}$\\ \hline
     			  
$\mathcal{M}^{\rm BC}_{\rm AAL}$ &$0.7648$ & $\!\!\!\begin{matrix} 0.9816, \\ (-0.1314,\\0.1314,\!0.6915) \end{matrix}$
     			 & $\!\!\begin{matrix} 0.9958, \\ (0.58,1.7241\\0.58) \end{matrix}$
     			  & $\begin{matrix} 0.9447, \\ (171.62,\\1.0117) \end{matrix}$ & $\!\!\!\!\begin{matrix} 0.9943, \\ (1.3262,\!1.0870,\\0.5796) \end{matrix}$
     			  & $\!\!\begin{matrix}  0.9981, \\ (3.92,0.70,\\0.36,0.7275) \end{matrix}$ & $\begin{matrix} 0.9942, \\ (10,2.2,0.60) \end{matrix}$ &$\!\!\!\!\begin{matrix} {0.9959}, \\ (0.640,\!1.6668,\\1.038,\!0.5641) \end{matrix}$ \\ \hline
     			  
$\mathcal{M}_{\rm AAL}$ &$0.1152$ & $\!\!\!\begin{matrix} 0.8586, \\ (-0.0346,\\0.0346,\!0.1484) \end{matrix}$
     			 & $\!\!\begin{matrix} 0.9432, \\ (0.14,7.1429\\0.14) \end{matrix}$
     			  & ${\rm Not~Valid}$ & $\!\!\!\begin{matrix} 0.9941, \\ (3.096,\!1.1183,\\0.1248) \end{matrix}$
     			  & $\!\!\begin{matrix}  0.9922, \\ (1.48,2.48,\\1.00,0.1232) \end{matrix}$ & $\begin{matrix} 0.9383, \\ (7.58,170.10,\\0.1386) \end{matrix}$ & $\!\!\!\begin{matrix} {0.9944}, \\ (1.075,\!3.2048,\\2.9222,\!0.1246) \end{matrix}$ \\ \hline
     			  
$\mathcal{M}_{\rm AAL}$ &$0.4339$ & $\!\!\!\begin{matrix}  0.2024, \\ (-0.2046,\\0.2046,\!1.2669) \end{matrix}$
     			 & $\!\!\begin{matrix} 0.4117, \\ (0.67,1.4925\\0.67) \end{matrix}$
     			  & $\begin{matrix} 0.1489, \\ (171.62,\\1.0117) \end{matrix}$ & $\!\!\!\begin{matrix} 0.4208, \\ (1.286,\!1.0803,\\0.6141) \end{matrix}$
     			  & $\!\!\begin{matrix}  0.4618, \\ (2.50,0.72,\\0.56,0.9489) \end{matrix}$ & $\begin{matrix} 0.3725, \\ (5,1.92,0.825) \end{matrix}$ & $\!\!\!\begin{matrix} {0.5903}, \\ (2.0271,\!0.988,\\73.720,\!0.3390) \end{matrix}$\\ \hline
     			  
$\mathcal{M}_{\rm AAL}$ &$0.8862$ & $\!\!\!\begin{matrix}  0.9439, \\ (-0.1229,\\0.1229,\!0.6349) \end{matrix}$
     			 & $\!\!\begin{matrix}  0.9982, \\ (0.93,1.0753\\0.93) \end{matrix}$
     			  & $\begin{matrix} 0.9983, \\ (17.75,\\1.1127) \end{matrix}$ & $\!\!\!\begin{matrix} 0.9976, \\ (1.012,1.005,\\ 0.9765) \end{matrix}$
     			  & $\!\!\begin{matrix}  0.9958, \\ (2.50,0.70,\\0.50,1.0122) \end{matrix}$ & $\!\!\begin{matrix} 0.9972, \\ (5,1.18,1.2169) \end{matrix}$ &$\!\!\!\begin{matrix} {0.9996}, \\ (0.6302,\!1.178,\\0.8444,\!0.9732) \end{matrix}$ \\ \hline
\hline
\end{tabular}
\end{table*}

This subsection provides a comprehensive experimental study on the statistical distribution of turbulence-induced fading in UWOC channels under the temperature random variations. As explained in Section V, two different procedures, i.e., using tunable water droppers and heating elements, are adopted to create such random variations on the water temperature across the beam propagation path. Moreover, statistical studies on the distribution of the received optical intensity, when both air bubbles and random temperature variations exist, are performed for the sake of generality. All the experiments are performed when the receiver possess the AAL; however, the effect of the transmitter BC is yet investigated. To simplify the notation, throughout this subsection, we denote the UWOC link configuration as $\mathcal{X}_{\rm AAL}$ if the link only employs the receiver AAL and as $\mathcal{X}^{\rm BC}_{\rm AAL}$ if it possesses the transmitter BC as well. In this general notation, $\mathcal{X}$ is a symbol representing the channel condition, i.e., $\mathcal{X}=\mathcal{H}$, $\mathfrak{D}$, and $\mathcal{M}$ for the UWOC link under temperature random variations caused by heating elements, temperature random variations induced by tunable water droppers, and the mixed presence of air bubbles and temperature random variations, respectively.
For all of these scenarios we run lots of experiments to cover a wide range of scintillation index and channel conditions.
Samples of the received intensity in the temporal domain are shown in Figs. \ref{fig_four} and \ref{fig_six}, while the corresponding histograms of the normalized received optical power as well as fitted statistical distributions are depicted in Figs. \ref{fig_five} and \ref{fig_seven}. Furthermore, detailed simulation results for GoF values of each of the discussed statistical distributions in Sec. IV and their corresponding constant coefficients are listed in Tables II and III. 

 Figs. \ref{fig_four}(a)-(c) illustrate the received signal through a UWOC link under the temperature random variations caused by passing the optical beam through heating elements when either the transmitter BC is used or not used in conjunction with the receiver AAL, i.e., a $\mathcal{H}^{\rm BC}_{\rm AAL}$ or a $\mathcal{H}_{\rm AAL}$ link, respectively. 
 As these figures and the corresponding histograms of the acquired data in Figs. \ref{fig_five}(a)-(c) suggest, for such scenarios the channel fading is relatively weak; mainly because the fairly large AAL appropriately compensates for the beam wandering caused by the heating elements. As a consequence and as it is demonstrated by the detailed numerical results in the first-third rows of Table II, lots of the considered statistical distributions can aptly fit the experimental data. In particular, both the Weibull and generalized Gamma distributions can match the acquired experimental data with the excellent GoF values larger than $0.99$.
 
 Figs. \ref{fig_four}(d)-(i) show time-domain samples of the received optical signal from a UWOC link under the temperature random variations caused by tunable water droppers launching three independent flow of hot water, with the temperature $T_h=90^{\rm o}$ \si{C}, into the cold water of tank, with the temperature $T_c=20^{\rm o}$ \si{C}, when only the receiver AAL is used in the configuration, i.e., a $\mathfrak{D}_{\rm AAL}$ link.
 Elaborating the corresponding histograms of the acquired data in Figs. \ref{fig_five}(d)-(i) and the detailed numerical results provided in the fourth-ninth rows of Table II reveals that not all the statistical distributions considered in this paper are capable of modeling the statistical behavior of UWOC channels under the temperature random variations, especially when the channel is under weak to moderate fading regions, i.e., $\sigma^2_I\approx 0.1$ and below this value. For example, for the case of $\sigma^2_I=0.0577$, as it is shown in Fig. \ref{fig_five}(e) and the fifth row of Table II, only the generalized Gamma distribution can excellently predict the statistical behavior of the channel; although the Weibull and exponentiated Weibull can fairly match the acquired data, the other distributions like Gamma and lognormal fail to model the fading distribution. On the other hand, as the plots in Figs. \ref{fig_five}(g)-(i) and the numerical results in the seventh-ninth rows of Table II suggest, when the scintillation index is fairly above $0.1$, the other simple distributions such as Gamma and lognormal can also acceptably match the acquired experimental data. In summary, for the $\mathfrak{D}_{\rm AAL}$ links, the generalized Gamma yields an excellent match to the experimental data for all the regions of the scintillation index and can be regarded as a general statistical model. The Weibull and exponentiated Weibull also fairly predict the fading distribution in all the regions of turbulence, while the simpler distributions like lognormal and Gamma may fail in some situations, especially for the scintillation index values around $0.1$ and below that.
 
 In order to investigate the effect of the transmitter BC on the statistical distribution of the received optical signal through a turbulent UWOC channel under the temperature random variations caused by tunable water droppers, similar experiments to the previous part are carried out. Figs. \ref{fig_four}(j)-(l) illustrate samples of the received optical signal, in temporal domain, through a $\mathfrak{B}_{\rm AAL}^{\rm BC}$ link. Carefully analyzing the corresponding histograms of the acquired experimental data in Figs. \ref{fig_five}(j)-(l) reveals that 
 using the transmitter BC not only significantly reduces the fading strength and alleviates the fading effect (in the cost of the reduced received optical power since the beam is expanded in the spatial domain) but also causes the received intensity samples to concentrate around the mean of the acquired data. As a consequence, and as the numerical results of the tenth-twelfth rows of Table II suggest, even though the scintillation index is around $0.1$ and below that, relatively all the statistical distributions can aptly model the fading statistics; however, once again, the generalized Gamma distribution provides the best fit to the acquired data.
 
 In the remaining part of this subsection, we deal with the statistical investigation of fading in turbulent UWOC channels under the mixed presence of air bubbles and random temperature variations. Fig. \ref{fig_six} shows samples of the received optical signal in the temporal domain through such a more general turbulent UWOC channel when either only the receiver AAL is used or also the transmitter BC is employed as well, i.e., either a $\mathcal{M}_{\rm AAL}$ link or a $\mathcal{M}^{\rm BC}_{\rm AAL}$ link. Excavating the results in Fig. \ref{fig_seven} for the histograms of the acquired experimental data and the corresponding detailed numerical results in Table III reveals that, as expected, the channel statistical behavior relatively contains the features of all $\mathfrak{B}_{(.)}^{(.)}$, $\mathcal{H}_{(.)}^{(.)}$, and $\mathfrak{D}_{(.)}^{(.)}$ links. For example, analyzing the results in Figs. \ref{fig_seven}(d)-(f) for $\mathcal{M}_{\rm AAL}$ links demonstrates that when the scintillation index is around ``$0.1$", e.g., Fig. \ref{fig_seven}(d), the channel statistical behavior is similar to that of $\mathfrak{D}_{\rm AAL}$ links, e.g., Figs. \ref{fig_five}(e) and (f) where not all the considered statistical distributions in this paper give the excellent match to the experimental data and the generalized Gamma and (exponential) Weibull distributions are the best candidates. More importantly, when the effect of air bubbles dominates and the channel is in moderate-to-strong fading region, like Fig. \ref{fig_seven}(e), the channel behavior becomes similar to that of $\mathfrak{B}_{\rm AAL}$ links or even $\mathfrak{B}$ links, like Figs. \ref{fig_three}(c), (d), and (f), where the single-lobe statistical distributions considered in this paper are no longer capable of modeling the fading statistics; in such cases, as investigated in our previous research \cite{jamali2016statistical}, general two-lobe statistical models which effectively combine two different distributions together, such as the mixed exponential-lognormal \cite{jamali2016statistical}, are required to better model the channel fading behavior.
 Similarly, exploring Figs. \ref{fig_seven}(a)-(c) and the first-third rows of Table III for the $\mathcal{M}^{\rm BC}_{\rm AAL}$ links shows that employing the BC at the transmitter side in conjunction with the receiver AAL, in addition to reducing the scintillation index and alleviating the fading deteriorating effect (in the cost of the reduced power reaching the receiver), decreases the configuration sensitivity to beam scattering. Therefore, the histogram of the acquired experimental data  becomes more smooth and no longer obeys a two-lobe behavior; consequently, the relatively simple single-lobe statistical distributions considered in this paper can excellently model the fading distribution. 
 
  \subsection{Turbulence-Induced Fading due to the Salinity Random Variations}
  \begin{figure*}
      \centering
      \includegraphics[trim=0cm 0cm 0cm 0cm,width=7in,height=3.2in,clip]{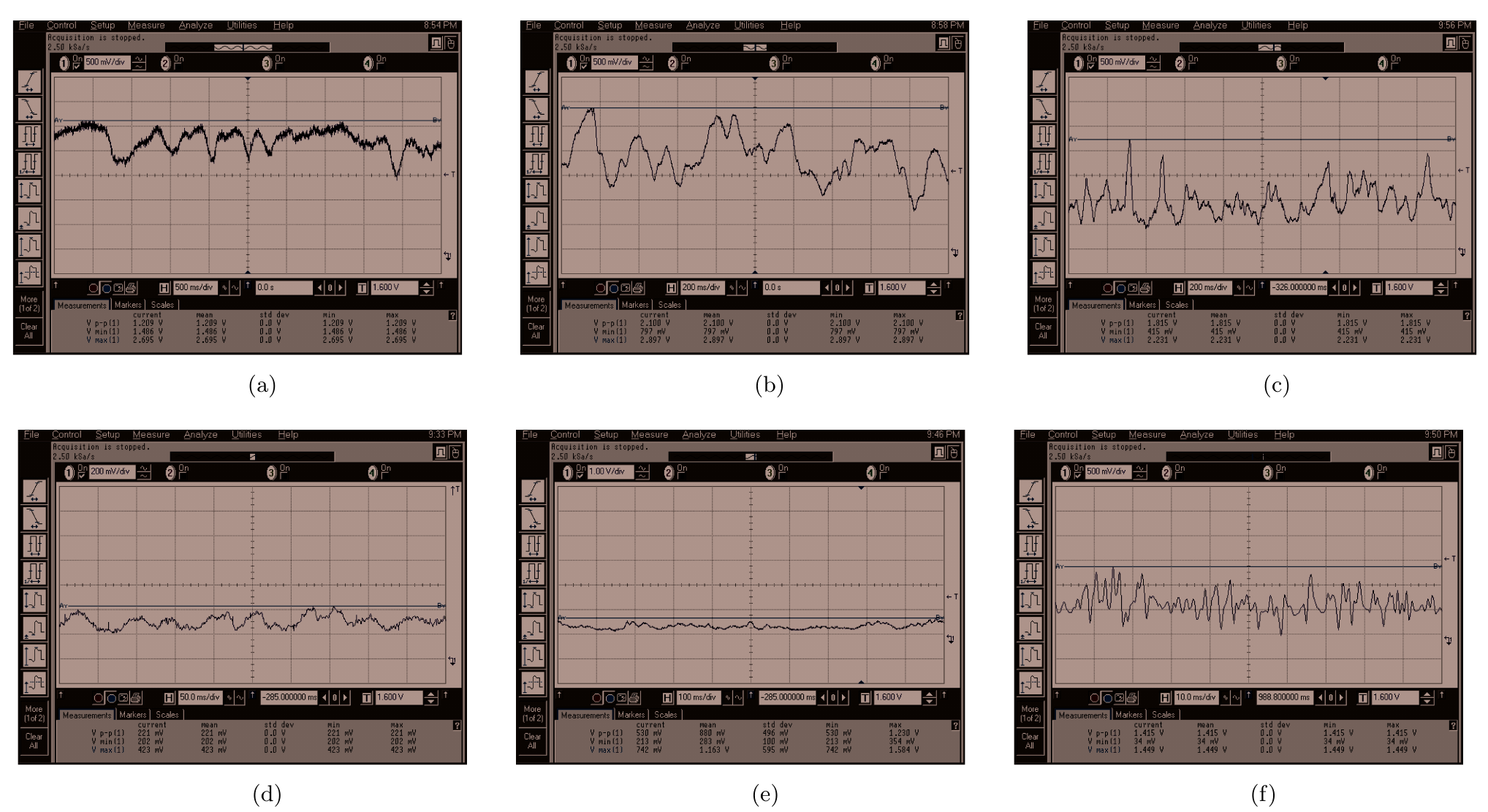}
      \caption{Received optical signal through (a) a UWOC link, under the salinity random variations, equipped with both the transmitter BC and the receiver AAL, $\mathcal{S}^{\rm BC}_{\rm AAL}$, with $\sigma^2_I=0.0092$; (b) a $\mathcal{S}^{\rm BC}_{\rm AAL}$ link with $\sigma^2_I=0.0546$; (c) a $\mathcal{S}^{\rm BC}_{\rm AAL}$ link with $\sigma^2_I=0.1181$; (d) a $\mathcal{S}^{\rm BC}_{\rm AAL}$ link with $\sigma^2_I=0.2023$; (e) a $\mathcal{S}^{\rm BC}_{\rm AAL}$ link with $\sigma^2_I=0.2783$; and (f) a $\mathcal{S}^{\rm BC}_{\rm AAL}$ link with $\sigma^2_I=0.3709$.}
      \vspace{-0.15in}
      \label{fig_eight}
      \end{figure*}
      
\begin{figure*}
      \centering
      \includegraphics[trim=0cm 0cm 0cm 0cm,width=7in,height=3.1in,clip]{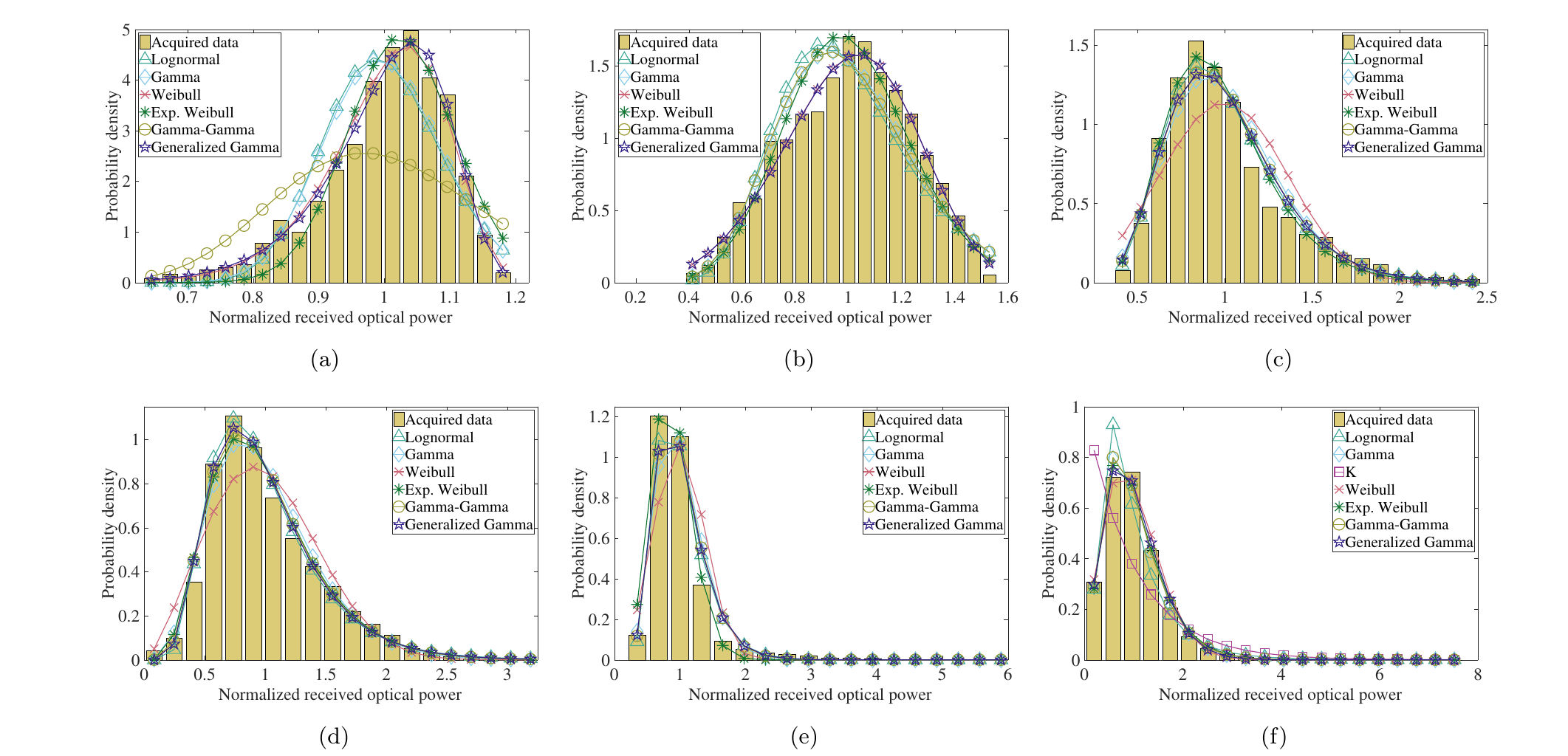}
      \caption{Accordance of different statistical distributions, considered in Sec. IV, with the histograms of the acquired data through various UWOC channel conditions under the salinity random variations, namely  (a) a UWOC link, under the salinity random variations, equipped with both the transmitter BC and the receiver AAL, $\mathcal{S}^{\rm BC}_{\rm AAL}$, with $\sigma^2_I=0.0092$; (b) a $\mathcal{S}^{\rm BC}_{\rm AAL}$ link with $\sigma^2_I=0.0546$; (c) a $\mathcal{S}^{\rm BC}_{\rm AAL}$ link with $\sigma^2_I=0.1181$; (d) a $\mathcal{S}^{\rm BC}_{\rm AAL}$ link with $\sigma^2_I=0.2023$; (e) a $\mathcal{S}^{\rm BC}_{\rm AAL}$ link with $\sigma^2_I=0.2783$; and (f) a $\mathcal{S}^{\rm BC}_{\rm AAL}$ link with $\sigma^2_I=0.3709$.}
      \label{fig_nine}
      \end{figure*}
      
\begin{figure*}
      \centering
      \includegraphics[trim=1cm 0cm 0cm 0cm,width=7.5in,clip]{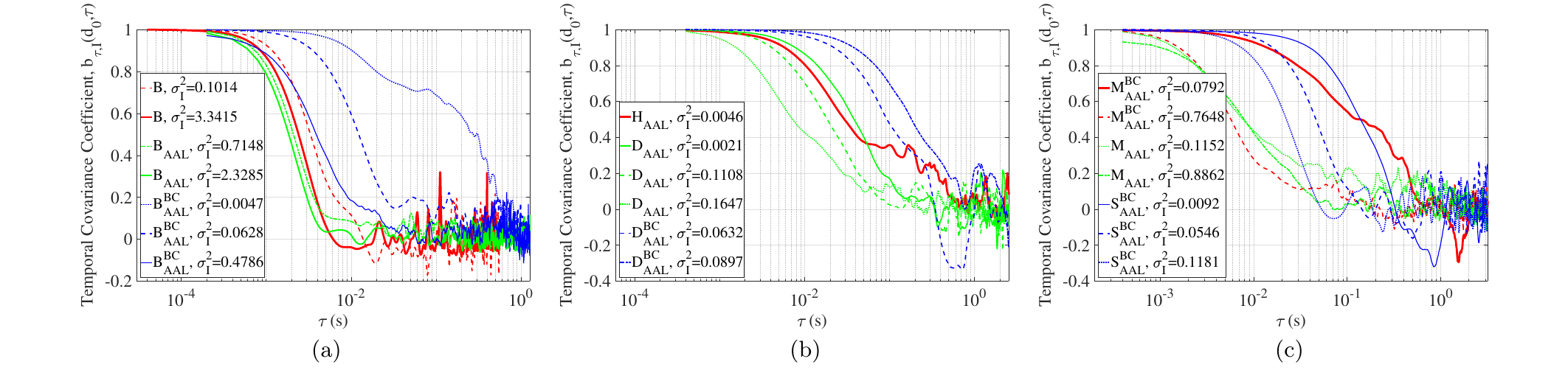}
      \caption{Temporal covariance coefficient, $b_{\tau,I}(d_0,\tau)$,
      for various channel scenarios.}
      \vspace{-0.05in}
      \label{fig_ten}
\end{figure*}
      
\begin{table*}
     		\centering
     		\caption{GoF and the constant parameters of different PDFs for the various channel conditions considered in Scenario III (Sec. VI-C).}
     		\begin{tabular}{||p{0.45in}||p{0.35in}||p{0.6in}||p{0.6in}||p{0.6in}||p{0.6in}||p{0.6in}||p{0.7in}||p{0.6in}||}
     			\hline\hline
     			\! Channel condition \vspace{-2cm} & \vspace{0.25cm} $\sigma^2_{I,{m}}$& 
    \vspace{0.1cm} {Log-normal}&
    \vspace{0.1cm} {Gamma}&
    \vspace{0.1cm} {K dist.}&
    \vspace{0.1cm} {Weibull}&
    \vspace{0.1cm} $\!\!${Exp. Weibull}&
    \vspace{0.1cm} $\!\!${Gamma-Gamma}&
    \vspace{0.1cm} $\!\!${Generalized Gamma}\\
     			\cline{3-9}
    && $\begin{matrix} \!{\rm GoF}, \\ \!(\mu_X,\sigma^2_X,\sigma^2_I) \end{matrix}$ &
     $\begin{matrix} {\rm GoF}, \\ (\theta,k,\sigma^2_I) \end{matrix}$ & 
     $\begin{matrix} {\rm GoF}, \\ (\alpha,\sigma^2_I) \end{matrix}$ & 
    $\begin{matrix} {\rm GoF}, \\ (\beta,\eta,\sigma^2_I) \end{matrix}$ &
    $\begin{matrix} {\rm GoF}, \\ (\alpha,\beta,\eta,\sigma^2_I) \end{matrix}$ & 
     $\begin{matrix} {\rm GoF}, \\ (\alpha,\beta,\sigma^2_I) \end{matrix}$ & 
    $\begin{matrix} {\rm GoF}, \\ (a,d,p,\sigma^2_I) \end{matrix}$ 
    \\ \hline \hline 			
$\mathcal{S}^{\rm BC}_{\rm AAL}$ &$0.0092$ & $\!\!\!\begin{matrix} 0.8015, \\ (-0.0021,\\0.0021,\!0.0082) \end{matrix}$
& $\!\!\!\!\begin{matrix} 0.8261, \\ (0.0083,\!121.21\\0.0083) \end{matrix}$
& ${\rm Not~Valid}$ & $\!\!\!\!\begin{matrix} 0.9805, \\ (13.162,\!1.0401,\\0.0086) \end{matrix}$
& $\!\!\!\!\begin{matrix} 0.9416, \\ (7.280,5.695,\\0.880,0.0065) \end{matrix}$ & $\begin{matrix} 0.5177, \\ (79.53,79.52,\\0.0253) \end{matrix}$ & $\!\!\!\begin{matrix} {0.9864}, \\ (1.072,\!11.610,\\16.879,\!0.0115) \end{matrix}$ \\ \hline

$\mathcal{S}^{\rm BC}_{\rm AAL}$ &$0.0546$ & $\!\!\!\begin{matrix} 0.7835, \\ (-0.0167,\\0.0167,\!0.0693) \end{matrix}$
& $\!\!\!\!\begin{matrix} 0.8671, \\ (0.0661,\!15.128\\0.0661) \end{matrix}$
& ${\rm Not~Valid}$ & $\!\!\!\!\begin{matrix} 0.9738, \\ (4.579,\!1.0947,\\0.0616) \end{matrix}$
& $\!\!\!\!\begin{matrix} 0.9033, \\ (3.091,2.694,\\0.830,0.0546) \end{matrix}$ & $\begin{matrix} 0.8544, \\ (155,16.5,\\0.0674) \end{matrix}$ & $\!\!\!\begin{matrix} {0.9738}, \\ (1.099,\!4.555,\\4.6208,\!0.0695) \end{matrix}$\\ \hline

$\mathcal{S}^{\rm BC}_{\rm AAL}$ &$0.1181$ & $\!\!\!\begin{matrix} 0.9745, \\ (-0.026,\\0.026,\!0.1095) \end{matrix}$
& $\!\!\begin{matrix} 0.9373, \\ (0.102,\!9.8039\\0.102) \end{matrix}$
& ${\rm Not~Valid}$ & $\!\!\begin{matrix} 0.7996, \\ (3.271,1.115,\\0.1131) \end{matrix}$
& $\!\!\begin{matrix} 0.9791, \\ (9.12,1.36,\\0.46,0.1002
) \end{matrix}$ & $\begin{matrix} 0.9583, \\ (19.5,19.5,\\0.1052) \end{matrix}$ & $\!\!\!\begin{matrix} {0.9590}, \\ (3.808\!\times\!10^{-4},\\19.8200,\\0.4754,\!0.1072) \end{matrix}$ \\ \hline

$\mathcal{S}^{\rm BC}_{\rm AAL}$ &$0.2023$ & $\!\!\!\begin{matrix} 0.9894, \\ (-0.049,\\0.049,\!0.2158) \end{matrix}$
& $\!\!\begin{matrix} 0.9745, \\ (0.192,\!5.2083\\0.192) \end{matrix}$
& ${\rm Not~Valid}$ & $\!\!\begin{matrix} 0.9081, \\ (2.431,1.127,\\0.1925) \end{matrix}$
& $\!\!\begin{matrix} 0.9827, \\ (4.549,1.198,\\0.528,0.2023
) \end{matrix}$ & $\begin{matrix} 0.9864, \\ (10.375,10.35,\\0.2023) \end{matrix}$ & $\!\!\!\begin{matrix} {0.9884}, \\ (7.882\!\times\!10^{-6},\\15.3200,\\0.3288,\!0.2004) \end{matrix}$\\ \hline

$\mathcal{S}^{\rm BC}_{\rm AAL}$ &$0.2783$ & $\!\!\!\begin{matrix} 0.9780, \\ (-0.034,\\0.034,0.1484) \end{matrix}$
& $\!\!\begin{matrix} 0.9474, \\ (0.131,\!7.6336\\0.131) \end{matrix}$
& ${\rm Not~Valid}$ & $\!\!\begin{matrix} 0.8559, \\ (3.017,1.119,\\0.1308) \end{matrix}$
& $\!\!\begin{matrix} 0.9873, \\ (2.00,2.22,\\0.78,0.1168
) \end{matrix}$ & $\begin{matrix} 0.9648, \\ (14.90,14.90,\\0.1387) \end{matrix}$ & $\!\!\!\begin{matrix} {0.9678}, \\ (6.611\!\times\!10^{-5},\\17.8200,\\0.3970,\!0.1462) \end{matrix}$ \\ \hline

$\mathcal{S}^{\rm BC}_{\rm AAL}$ &$0.3709$ & $\!\!\!\begin{matrix} 0.9342, \\ (-0.1057,\\0.1057,\!0.5261) \end{matrix}$
& $\!\!\begin{matrix} 0.9918, \\ (0.338,\!2.9586\\0.338) \end{matrix}$
& $\!\!\begin{matrix} 0.5679, \\ (171.62,\\1.0117) \end{matrix}$ & $\!\!\begin{matrix} 0.9918, \\ (1.939,\!1.1276,\\0.2889) \end{matrix}$
& $\!\!\begin{matrix} 0.9937, \\ (2.16,1.24,\\0.74,0.3379
) \end{matrix}$ & $\begin{matrix} 0.9911, \\ (169.4,3.0,\\0.3412
) \end{matrix}$ & $\!\!\!\begin{matrix} {0.9960}, \\ (0.7598,\!2.327,\\1.4353,\!0.3068) \end{matrix}$\\ \hline

\end{tabular}
\end{table*}
In addition to the presence of air bubbles and temperature random variations, the salinity random variations is also one of the major causes of fading in UWOC channels. In fact, the theoretical studies on the light scintillation in turbulent oceanic media \cite{korotkova2012light} have demonstrated that the scintillation index is a monotonically increasing function of a parameter, so-called $w$, which defines the relative strength of temperature and salinity fluctuations. It is known that $w$ in the oceanic media varies in the range $[-5,0]$, and as $w$ increases and approaches its maximum value, i.e., ``$0$", the salinity-induced optical turbulence becomes dominant compared to the temperature-induced turbulence \cite{nikishov2000spectrum}. Therefore, since the scintillation index is an increasing function of $w$, one may expect that the salinity random variation can even induce a stronger turbulence in UWOC channels compared to the temperature random variations. In the meantime, to the best of our knowledge, experimental studies on the statistics of fading in UWOC channels under the salinity random variations is completely missing in the literature.
  
  Inspired by the lack of the aforementioned critical study in the literature, in this subsection we deal with the statistical studies on the fading of UWOC channels under the salinity random variations. Before going over such an experimental study on the fading statistics, we made another experiment to observe the response of salty UWOC channels; however, we do not include the results here for the sake of brevity. In that experiment, we added $700$ \si{g} salt to our water tank and observed a significant loss on the received optical signal compared to that of fresh water link. Moreover, we have observed that solving salt in the water, for such a low link range of our water tank, does not induce a noticeable fluctuation on the received signal. We then used a water pump to circulate water within the tank. Our extra experiments indicated that as the velocity of moving medium increases, the received signal experiences faster and more severe fluctuations, mainly because as the water flow increases, particles within the water move more rapidly and this causes more randomness on the propagation path.
  
  In the next part of our experiments, in order to carefully model the fading distribution in UWOC channels under salinity random variations, as explained in Sec. V, we used three independent water droppers to enter three flow of extremely salty water into the fresh water of the tank. For the sake of brevity and space limitation, here, we only present the results for the link configurations that employ both the transmitter BC and the receiver AAL together, i.e., the $\mathcal{S}^{\rm BC}_{\rm AAL}$ link configuration. Fig. \ref{fig_eight} illustrates temporal domain samples of the received optical signal from a $\mathcal{S}^{\rm BC}_{\rm AAL}$ link for various values of the scintillation index. Carefully exploring the histograms of the acquired experimental data in Fig. \ref{fig_nine} and the corresponding numerical results provided in Table IV reveals that not all the statistical distributions considered in this paper are able to excellently match the experimental data, especially when the scintillation index value is small, like Figs. \ref{fig_nine}(a) and (b); however, both the generalized Gamma and exponentiated Weibull distributions can perfectly model the fading statistical behavior in UWOC channels under the salinity random variations for the whole range of scintillation index considered in our experiments.
  
{ At the end, we should highlight that when all the three effects exist together, the histogram tends to follow the statistical behavior of the most dominant one, e.g., if the air population is high, the histogram tends to that of $\mathfrak{B}$ links and so on. Moreover, we should emphasize that when the sensitivity of the link geometry to the beam wandering is very high (depending on the link length, water type, optical source employed, and the link configuration adopted in real underwater environments), the received intensity will tend to mainly lie either in large or small values (see for example Fig. 2(c)). In such circumstances the typical single-lobe distributions cannot appropriately fit the experimental data and generally a two-lobes statistical distribution, such as the mixed exponential-lognormal distribution we have proposed in \cite{jamali2016statistical}, is required to capture the statistical behavior of UWOC fading in all regions of the scintillation index. In this regard, mixing the more suitable distributions testified in this paper such as generalized Gamma and exponentiated Weibull distributions with the exponential distribution (or some other proper distributions) will be a promising solution.}   
  \subsection{Coherence Time Evaluation}
To complete our statistical studies over the fading behavior in turbulent UWOC channels, in this subsection we present the simulation results for the channel coherence time as an important metric in describing the channel fading. As explained, the channel coherence time determines the average time in which the channel fading coefficient can be considered constant, or equivalently, the temporal covariance coefficient of irradiance, defined in Eq. \eqref{NTC}, is above a certain threshold level. 

Fig. \ref{fig_ten} illustrates the temporal covariance coefficient of irradiance for different scenarios, including $\mathfrak{B}$, $\mathcal{H}$, $\mathfrak{D}$, $\mathcal{M}$ and $\mathcal{S}$ links with or without the transmitter BC and the receiver AAL. It is observed that the channel coherence time for all of the different scenarios is usually larger than $10^{-3}$ seconds. Considering the relatively large rates of data transmission in the state-of-the-art UWOC systems, such values of the channel coherence time strongly confirm that the channel fading is slow, in the temporal domain, meaning that the same fading coefficient can be considered over thousands up to millions of consecutive bits. Carefully exploring the results reveals that using the transmitter BC, through alleviating the sensitivity of the link to the beam scattering, results into a larger coherence time values implying a more slowly varying fading. On the other hand, one may observe that, for a given channel condition, the fading temporal variation slightly increases (the channel coherence time decreases) with the increase on the scintillation index value, i.e, the higher the value of $\sigma^2_I$ is, the less the value of $b_{\tau,I}(d_0,\tau)$ we have.
 \section{Conclusions and Future Directions}
In this paper, we provided a comprehensive experimental study on the statistics of fading in UWOC channels under various conditions. To do so, we intentionally induced turbulence across the propagation path, using different procedures, and investigated the accordance of different statistical distributions in modeling the fading behavior for various channel scenarios. We also explored the channel coherence time, as an important metric in describing the fading behavior, to comment on the slow nature of fading in UWOC channels. Moreover, the effect of the transmitter beam-collimator and the receiver aperture averaging lens, as two integral parts in the implementation of UWOC systems, have been investigated in the fluctuations of the received optical signal.
Our extensive experimental and simulation results demonstrated that when the channel is in moderate-to-strong fading and is dominantly affected by the presence of air bubbles, the single-lobe statistical distributions cannot acceptably model the fading behavior, and in general, a two-lobe distribution, like the mixed exponential-lognormal distribution is required to describe the fading statistics. In the meantime, we observed that employing the transmitter beam-collimator and/or the receiver aperture averaging lens, through alleviating the link sensitivity to beam scattering, significantly reduces the aforementioned problem and suits the single-lobe statistical distributions considered in this paper. We also run lots of experiments to model the fading statistical behavior when the optical turbulence is produced by the random variation of water temperature and salinity in a wide range of scintillation index, from weak to strong fading regions. Our thorough numerical and experimental studies revealed that, although the simple statistical distributions such as lognormal and Gamma can aptly match the histogram of the acquired data in many of the channel conditions, there are some specific scenarios in which such simple distributions fail to model the fading. In this circumstances, it is observed that the generalized Gamma and exponentiated Weibull are, respectively, the best candidates that can excellently fit the experimental data though in the cost of increased complexity. Moreover, our numerical results for the irradiance covariance coefficient  in various scenarios indicated that the UWOC channel coherence time is usually greater than $10^{-3}$ implying that the channel fading coefficient can be considered constant over thousands up to millions of consecutive bits.
{We should emphasize that although some of the considered well-known statistical distributions do not appropriately described the statistical behavior of fading under water, it is very important to determine how much they deviate from the actual measurement data depending on the specific channel condition and link configuration adopted; this will significantly shape the future research, especially over the performance evaluation of advanced UWOC systems, in an appropriate direction.}

{ At the end, we should emphasize that perfect understanding of UWOC systems requires the following two important phases: UWOC channel modeling, and UWOC performance analysis. In this regard, while this paper attempts to study the statistical distribution of fading under water through comprehensive experiments, the mathematical modeling of UWOC fading using well-known theorems in the statistical optics, like the Karhunen-Lo\'{e}ve expansion, to obtain the general statistical distributions suitable for predicting the fading behavior in turbulent UWOC channels can be considered as a rigorous and mature future research. In this case, the extensive experimental data provided in this paper can be regarded as good benchmarks for evaluating the accuracy of such mathematical models. On the other hand, using the state-of-the-art optical transmitter and receiver pairs in order to extend the current comprehensive laboratory-based statistical study to real sea environments and then investigating the fading behavior in such scenarios can be considered as a powerful and useful direction for the future research in this area. Moreover, after advancing the first phase corresponding to the UWOC channel modeling, the results from the UWOC statistics can be applied to theoretically analyze the performance of UWOC systems employing advanced transmission/reception techniques with respect to the accurate UWOC channel model and also experimentally evaluate the performance of such systems in well-developed experimental setups.}


%


\end{document}